\documentclass[preprint,3p,review,12pt]{elsarticle}

\usepackage[all]{nowidow}
\usepackage{comment,color}
\usepackage{amssymb} 
\usepackage{physics}
 
\makeatletter
\def\ps@pprintTitle{%
    \let\@oddhead\@empty
    \let\@evenhead\@empty
    \def\@oddfoot{\footnotesize\itshape
         {Submitted} \hfill\today}%
    \let\@evenfoot\@oddfoot
    }
\makeatother

\usepackage{amsmath}
\usepackage{lineno}

\newcommand*\patchAmsMathEnvironmentForLineno[1]{%
  \expandafter\let\csname old#1\expandafter\endcsname\csname #1\endcsname
  \expandafter\let\csname oldend#1\expandafter\endcsname\csname end#1\endcsname
  \renewenvironment{#1}%
     {\linenomath\csname old#1\endcsname}%
     {\csname oldend#1\endcsname\endlinenomath}}% 
\newcommand*\patchBothAmsMathEnvironmentsForLineno[1]{%
  \patchAmsMathEnvironmentForLineno{#1}%
  \patchAmsMathEnvironmentForLineno{#1*}}%
\AtBeginDocument{%
\patchBothAmsMathEnvironmentsForLineno{equation}%
\patchBothAmsMathEnvironmentsForLineno{align}%
\patchBothAmsMathEnvironmentsForLineno{flalign}%
\patchBothAmsMathEnvironmentsForLineno{alignat}%
\patchBothAmsMathEnvironmentsForLineno{gather}%
\patchBothAmsMathEnvironmentsForLineno{multline}%
}
\journal{ }
\setcitestyle{square}

\begin{document}

\begin{frontmatter}



\title{Limits to predictability of the asymptotic state of the Atlantic Meridional Overturning Circulation in a conceptual climate model}

% \Author[affil]{given_name}{surname}
\author[1]{Oliver Mehling\corref{cor1}}
\ead{oliver.mehling[at]polito.it}
\cortext[cor1]{Corresponding author}
\author[2,3]{Reyk Börner}
\author[2,3,4]{Valerio Lucarini}

\affiliation[1]{organization={DIATI – Department of Environment, Land and Infrastructure Engineering, Politecnico di Torino},
             addressline={Corso Duca degli Abruzzi 24},
             postcode={10129},
             city={Turin},
             country={Italy}}
\affiliation[2]{organization={Department of Mathematics and Statistics},
             addressline={University of Reading},
             city={Reading},
             postcode={RG6 6AX},
             country={United Kingdom}}
\affiliation[3]{organization={Centre for the Mathematics of Planet Earth},
             addressline={University of Reading},
             city={Reading},
             postcode={RG6 6AX},
             country={United Kingdom}}
             \affiliation[4]{organization={School of Mathematical and Computational Sciences},
             addressline={University of Leicester},
             city={Leicester},
             postcode={LE1 7RH},
             country={United Kingdom}}

%%%% Abstract %%%%
\begin{abstract}
Anticipating critical transitions in the Earth system is of great societal relevance, yet there may be intrinsic limitations to their predictability. For instance, from the theory of dynamical systems possessing multiple chaotic attractors, it is known that the asymptotic state depends sensitively on the initial condition in the proximity of a fractal basin boundary. Here, we approach the problem of final-state sensitivity of the Atlantic Meridional Overturning Circulation (AMOC) using a conceptual climate model, composed of a slow bistable ocean coupled to a fast chaotic atmosphere. First, we explore the occurrence of long chaotic transients in the monostable regime, which can mask a loss of stability near bifurcations. In the bistable regime, we explicitly construct the chaotic saddle using the edge tracking technique. Quantifying the final-state sensitivity through the maximum Lyapunov exponent and the lifetime of the saddle, we find that the system exhibits a fractal basin boundary with almost full phase space dimension, implying vanishing predictability of the second kind near the basin boundary. Our results demonstrate the usefulness of studying non-attracting chaotic sets in the context of predicting climatic tipping points, and provide guidance for the interpretation of higher-dimensional models such as general circulation models.
\end{abstract}


\end{frontmatter}


%\begin{linenumbers}
\section{Introduction}
Like many systems in nature, several elements of the Earth system are thought to be multistable: for a given climatic forcing, they may possess multiple competing attractors that can be reached from different initial conditions \citep{Feudel2018}. Multistability has been demonstrated in rather complex physical models of the Greenland \citep{Robinson2012} and West Antarctic \citep{Garbe2020} ice sheets, the Amazon rainforest \citep{Oyama2003}, the Atlantic Meridional Overturning Circulation (AMOC) \citep{Hawkins2011}, and even Earth as a whole \citep{Margazoglou2021}, supported by paleoclimatic evidence for abrupt climate changes in the past \citep{Brovkin2021,Boers2022}. The proposed multistability of the AMOC  has been intensively studied for several decades \citep{Dijkstra2005a,Kuhlbrodt2007,Weijer2019} and has recently attracted renewed attention due to a suggested loss of stability over the past century seen in observation-based indicators \citep{Boers2021}. The paradigm of a multistable AMOC dates back to the seminal model introduced by Stommel \citep{Stommel1961}, who showed that there can be two competing (``on'' and ``off'') states for a given freshwater forcing due to the positive salt-advection feedback. This concept has been invoked to explain qualitative changes of the ocean circulation in past climates \citep{Broecker1985,Rahmstorf2002} as well as the response of climate models to changes in the hydrological cycle \citep{Rahmstorf1995,Stocker1997,Rahmstorf2005}.

Collectively, the multistable climate subsystems mentioned above are subsumed under the term tipping elements \citep{Lenton2008}, as anthropogenic global warming may trigger a possibly irreversible transition (tipping) to a qualitatively different state on timescales relevant for policy \citep{ArmstrongMcKay2022}. In turn, the concept of a ``safe operating space'' has been framed as keeping a control parameter (such as greenhouse gas forcing) below a critical value to avoid the transition into an undesired state of the Earth system \citep{Rockstrom2009}. However, in addition to the classical picture of crossing a bifurcation (``bifurcation tipping''), internal variability and the rate of the forcing can also lead to noise-induced and rate-induced tipping, respectively \citep{Ashwin2012}, which means that in reality the safe operating space is likely more complex than what can be described with a single critical value \citep{Lucarini2007,Alkhayuon2019,Lohmann2021}.

By incorporating the dominant physical processes into low-dimensional conceptual models, dynamical systems theory has proven extremely useful in advancing our understanding of the mechanisms of tipping points in the Earth system (see \citep{Boers2022,Kuehn2011,Ghil2020,Ritchie2023,Dijkstra2024} for recent reviews on this topic). For example, the derivation of statistical early warning indicators \citep{Dakos2012,Lenton2011,SantosGutierrez2022}, which can be applied to observational (e.g., \citep{Boers2021,Ditlevsen2023}) or modeled (e.g., \citep{Boulton2014}) climatic time series, has increased the prospect of anticipating that a bifurcation is approached. However, in a multistable chaotic system there are fundamental limitations to predictability of the final state even in the autonomous case. The question of whether the attractor reached from a given initial condition can be accurately predicted has been coined ``predictability of the second kind'' by Lorenz \citep{Lorenz1975}. This is in contrast to predictability ``of the first kind'', which refers to the ability to predict the future state of a system at a given horizon, given the knowledge of the initial conditions with finite precision. The presence of a limited predictability of the first kind is a key characteristic of chaotic systems and is usually investigated through Lyapunov analysis \cite{kalnay2003,Pikovsky2016}.  

Let us consider the simpler case of bistable systems. Uncertainty on our ability to predict the final state emerges from the (unavoidably) finite observation time. Beyond a bifurcation point, one may encounter so-called ``ghost attractors'' – states that are not asymptotically stable but feature transient chaos with finite yet possibly very long lifetimes, which depend sensitively on the initial conditions \citep{Lai2011}. In practice, this means that the system appears to reside in a well-defined steady state until a sudden transition to the actual attractor occurs. This phenomenon is common in many areas of physics; for example, turbulence in pipe flows is often regarded as a very long chaotic transient rather than a genuine attractor (e.g., \citep{Tel2008}). In this case, given a finite time of observation or simulation (such as in state-of-the-art climate models, which are typically integrated for decades to centuries), one might misjudge the stability properties of the system. This is particularly relevant when a control parameter is close to the critical value separating monostable and bistable behavior.

In the parametric region where bistability is observed, a separate instance of difficulty emerges in predicting the asymptotic state from the choice of the initial condition. In a chaotic, bistable $D$-dimensional system, limited final state predictability is directly linked to the presence of a fractal boundary between the two basins of attraction with fractal dimension $D - 1 \leq D_b < D$ \citep{McDonald1985}: given that an initial condition $\mathbf{u}_0$ can only be determined to a precision $\varepsilon$, the fraction $f$ of (a bounded region of) phase space in which the outcome is uncertain (i.e., different attractors can be reached from within $\mathbf{u}_0 \pm \varepsilon$) scales like
\begin{align}
f \propto \varepsilon^\alpha, \label{eq:predictability}
\end{align}
where $\alpha = D - D_b$ is the uncertainty exponent \citep{McDonald1985}. In practical terms, $\alpha \ll 1$ means that decreasing the uncertainty $\varepsilon$ only yields a very small improvement in final state predictability as given by $f$, and the phase space region around the boundary is essentially a ``grey zone'' in which the final state is almost unpredictable. The case $\alpha \ll 1$ is believed to be relatively common \citep{Grebogi1987}. It has been conjectured that the fractal basin boundary dimension is linked to the Lyapunov spectrum and the lifetime of the chaotic saddle \citep{Hunt1996,Sweet2000}, with $\alpha \ll 1$ being associated with a chaotic instability on the saddle that is fast compared to its lifetime \citep{Bodai2020}. While their importance for the transient and asymptotic behavior of chaotic dynamical systems has long been recognized \citep{Lai2011,Ott2002}, both fractal basin boundaries and chaotic saddles remain understudied in the context of climatic tipping points. To our knowledge, the only study that explicitly determined the basin boundary dimension in the context of two competing climatic attractors has been by Lucarini \& Bódai \citep{Lucarini2017}. They used a climate model of intermediate complexity to investigate the ``Snowball Earth'' transition \citep{Budyko1969,Sellers1969}, in which the ice-albedo feedback drives almost every initial condition either to a fully glaciated or an ice-free climate, and found that the basin boundary has almost full dimension, with $\alpha \approx 0.02$. Regarding the AMOC, Lohmann \& Ditlevsen \citep{Lohmann2021} found that in the context of rate-induced tipping (occurring before the bifurcation point is crossed \citep{Ashwin2012}), the final AMOC state in a simplified ocean general circulation model (GCM) depends sensitively on the initial condition. They qualitatively linked this behavior to the presence of a fractal basin boundary but did not assess its properties explicitly.

In this study, we quantitatively explore the limits of predictability of the final AMOC state in a conceptual atmosphere--ocean model inspired by Gottwald \citep{Gottwald2021} (Sec. \ref{sec:model}). The model mimics a key feature of more complex GCMs: it exhibits chaotic dynamics driven by a fast atmospheric component, while the oceanic component acts as a slow integrator. The oceanic component introduces the bistability in the system as a result of the coexistence of the AMOC ``on'' and ``off'' states. We investigate aspects related to both time and phase space that limit predictability of the second kind. First, we focus on the lifetime of chaotic transients which may effectively prevent the system from tipping on finite timescales, even when overshooting the bifurcation (Section \ref{sec:transients}). Then, building on the methods of \citep{Lucarini2017}, we use the edge tracking algorithm \citep{Skufca2006} to construct the chaotic saddle (also called \textit{Melancholia state} by \citep{Lucarini2017}) between two competing AMOC states, whose lifetime and Lyapunov exponents are directly linked to the dimension of the fractal basin boundary \citep{Hunt1996,Sweet2000} (Section \ref{sec:saddle}). Finally, we discuss the implications of both phenomena on the predictability of AMOC tipping and on the concept of a safe operating space (Section \ref{sec:discussion}) and summarize our conclusions in Section \ref{sec:conclusions}.

\section{The conceptual atmosphere-ocean model} \label{sec:model}
First, we introduce the conceptual climate model which isolates the principal characteristics of atmosphere--ocean flow that we will focus on in this study -- a fast, chaotic atmosphere coupled to a slow, bistable ocean component. To this end, we use a coupled configuration of the seminal conceptual models for mid-latitude atmospheric flow and the AMOC, respectively: the Lorenz-84 model \citep{Lorenz1984} (L84 hereafter) and the Stommel model \citep{Stommel1961}.

\subsection{Model components}
\begin{figure}[htbp]
    \centering
    \includegraphics[width=0.5\textwidth]{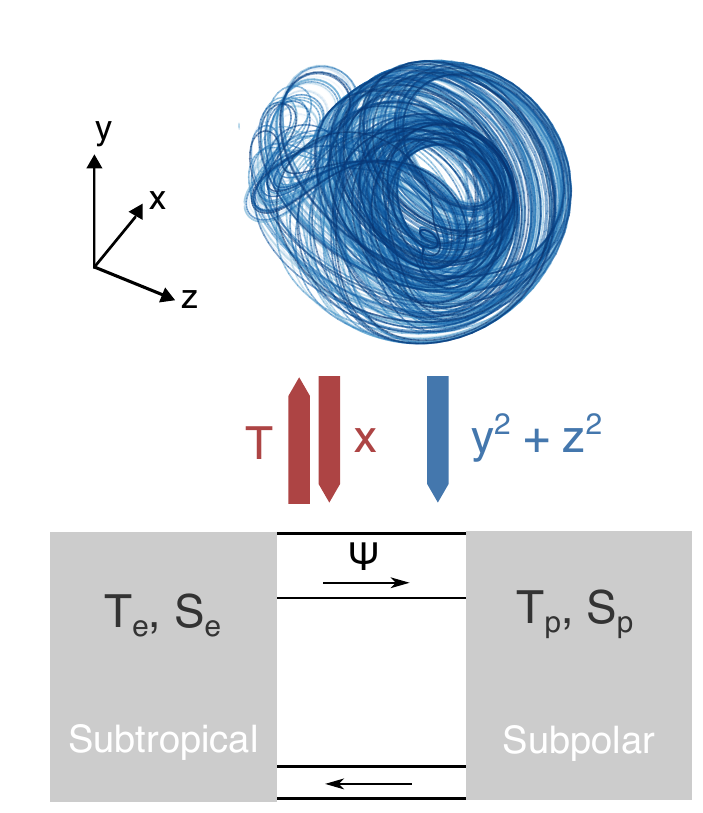}
    \caption{Sketch of the coupled L84-Stommel model. In the Stommel model, $T_e$ and $S_e$ are temperature and salinity in the subtropical North Atlantic box, and $T_p$ and $S_p$ are temperature and salinity in the subpolar North Atlantic box, with $T \equiv T_e-T_p$ and $S \equiv S_e-S_p$. The atmosphere--ocean coupling terms via heat fluxes and freshwater fluxes are sketched in red and blue, respectively.}
    \label{fig:model-sketch}
\end{figure}

The L84 model for the atmospheric mid-latitude circulation is given by a set of three ordinary differential equations \citep{Lorenz1984}:
\begin{align} \label{eq:L84}
\begin{split}
\dot{x} &= -(y^2 + z^2) - a\,(x - F) \\
\dot{y} &= x y - b x z - (y - G) \\
\dot{z} &= b x y + x z - z,
\end{split}
\end{align}
where $x$ is the strength of the westerly winds (a conceptual representation of the jet stream) given by thermal wind balance, and $y$ and $z$ are the amplitudes of the cosine and sine phases of a superimposed traveling wave. The parameters $F$ and $G$ are the background values for the meridional and zonal temperature gradients, respectively, to which $x$ and $y$ would relax in an uncoupled setting; $a$ and $b$ control the internal timescales of the model and will be kept at their original values $a=0.25$ and $b=4$ \citep{Lorenz1984}. Despite its very simple representation of atmospheric flow, the model has a strong physical basis, as it can be derived by Galerkin truncation of a two-layer quasi-geostrophic model \citep{VanVeen2003}. It also exhibits a complicated bifurcation structure on its own \citep{Broer2002,Wang2014}, but since different states of atmospheric flow are not the focus of our study, we remain in the regime $G=1$ where the L84 model only has a single chaotic attractor depicted in Fig. \ref{fig:model-sketch}.

\begin{figure}[htbp]
    \centering
    \includegraphics[width=0.5\textwidth]{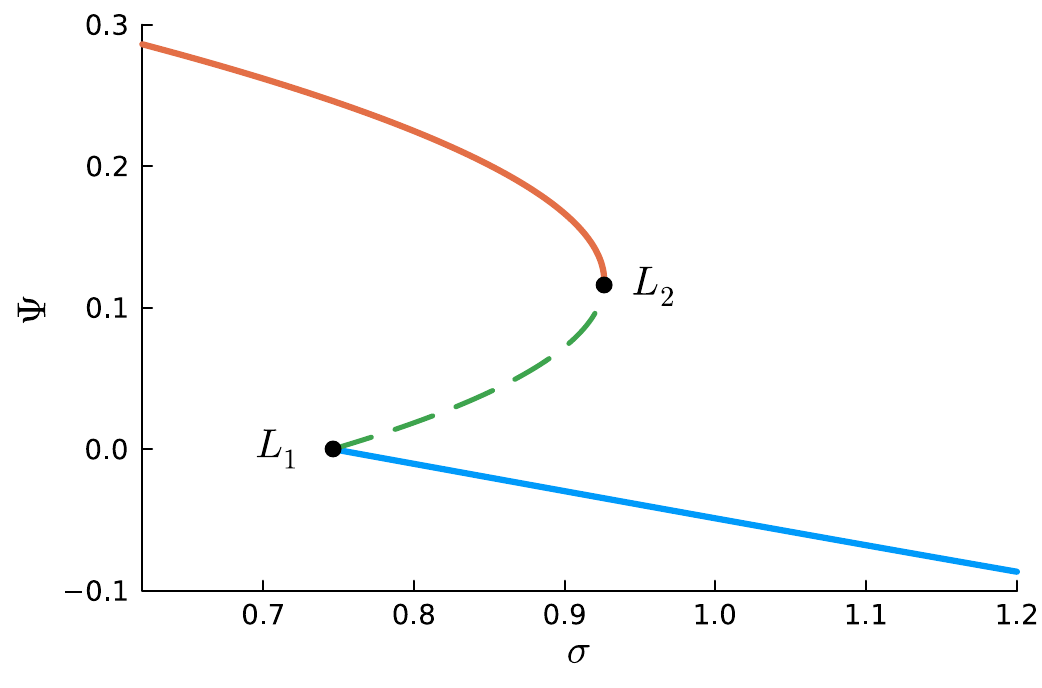}
    \caption{Bifurcation diagram of the uncoupled Stommel model (Eq. \ref{eq:stommel}): fixed points of the AMOC strength $\Psi$ as a function of the control parameter $\sigma$. The saddle-node bifurcation points are labeled as $L_1$ and $L_2$.}
    \label{fig:stommel-bifurcation}
\end{figure}

The Stommel model \citep{Stommel1961} is the canonical conceptual model for the bistable AMOC and describes the thermohaline flow in the North Atlantic as a function of the meridional temperature gradient $T$ and the salinity gradient $S$ between two boxes representing the subtropical and subpolar North Atlantic. For several decades, Stommel's model has served as a key model to interpret qualitative changes in the AMOC \citep{Rahmstorf1995,Kuhlbrodt2007,Weijer2019}. Here, we use the version derived by Cessi \citep{Cessi1994} and follow the notation of Gottwald \citep{Gottwald2021}:
\begin{align} \label{eq:stommel}
\begin{split}
\dot{T} &= -\frac{1}{\varepsilon_a}(T - \theta) - T - \mu |\Psi| T \\
\dot{S} &= \sigma - S - \mu |\Psi| S,
\end{split}
\end{align}
where $\varepsilon_a = t_r/t_d$ is the ratio between the restoring timescale for temperature $t_r$ and the diffusive timescale of the ocean $t_d$, $\mu = t_{ad}/t_d$ is the ratio between the advective timescale $t_{ad}$ and the diffusive timescale, $\theta$ is the surface temperature gradient to which $T$ is restored, and $\sigma$ is the surface freshwater flux. The advective terms are proportional to the absolute value of the AMOC strength $\Psi = T - S$, where $\Psi > 0$ represents a thermally driven circulation like in the present-day climate, and $\Psi < 0$ represents a salinity-driven AMOC where the circulation would be reversed. Here, as in many previous studies, we consider the freshwater flux $\sigma$ as the main control parameter that determines the AMOC regime. The corresponding bifurcation diagram, which reveals a bistable region bounded by two saddle-node bifurcations, is shown in Fig. \ref{fig:stommel-bifurcation}.

\subsection{Coupled model equations}
As in several previous studies \citep{Gottwald2021,Roebber1995,VanVeen2001} of conceptual atmosphere--ocean coupling, we couple the Stommel model to the L84 atmosphere through the temperature gradients and the freshwater flux (Fig. \ref{fig:model-sketch}). Here, we follow the recent formulation of coupling terms by Gottwald \citep{Gottwald2021} (without the sea ice component), who introduced an explicit timescale separation $\varepsilon_f$ between the atmospheric and oceanic components, similar to earlier studies \citep{Roebber1995,VanVeen2001}. The equations of the five-dimensional coupled model used in our study are given by:

\begin{align} \label{eq:coupled}
\begin{split}
\varepsilon_f \dot{x} &= -\Delta - a\,(x - F_0 - F_1 T) \\
\varepsilon_f \dot{y} &= x y - b x z - (y - G_0) \\
\varepsilon_f \dot{z} &= b x y + x z - z \\
\dot{T} &= -\frac{1}{\varepsilon_a}(T - T_\text{surf}) - T - \mu |S-T| T \\
\dot{S} &= S_\text{surf} - S - \mu |S-T| S
\end{split}
\end{align}

with

\begin{align*}
\Delta &= y^2 + z^2 \\
T_\text{surf} &= \theta_0 + \theta_1 \frac{x - \bar{x}}{\sqrt{\varepsilon_f}} \\
S_\text{surf} &= \sigma_0 + \sigma_1 \frac{\Delta - \bar{\Delta}}{\sqrt{\varepsilon_f}}
\end{align*}

Here, in a similar spirit as in \citep{Roebber1995,VanVeen2001}, we use a two-way coupling: the fast atmospheric component drives the ocean via the restoring temperature and salinity/freshwater flux, while the slow ocean component couples to the atmosphere via the meridional temperature gradient $F$. For simplicity, we keep the zonal temperature gradient $G=G_0$ constant, noting that the chaotic properties of the L84 model as characterized by its maximum Lyapunov exponent already follow an intricate non-monotonic structure when only $F$ is varied \citep{Freire2008}.

The main coupling constant is $F_1$, which we set to $F_1 = 0.1$ following the similar model of \citep{VanVeen2001}, who argued that the atmosphere should be weakly coupled to the ocean. Furthermore, the timescale separation parameter between atmosphere and ocean is chosen as $\varepsilon_f = 3 \cdot 10^{-4}$, which corresponds to a timescale of about 10 days for the atmosphere and 100 years for the ocean. Note that this oceanic timescale defines the scale of the dimensionless time $t$, such that all times given here can be read in a ``unit'' of centuries. Our oceanic timescale is slightly shorter than the one given by Cessi \citep{Cessi1994}, but we mostly aim at a realistic order of magnitude, as a smaller $\varepsilon_f$ needs to be traded off against a higher computational cost for the edge tracking. Except for $F_1$ and $\varepsilon_f$, we use the default parameters of \citep{Gottwald2021} (see \ref{appendix:model-parameters} for a full list of parameter definitions and values). The equations are solved numerically using a fifth-order Runge-Kutta scheme with a fixed timestep of $7.5 \cdot 10^{-6}$, which corresponds to a timestep of $1/40$ of the atmospheric timescale as used in \citep{Roebber1995}. 

We remark that it is possible to perform a rigorous asymptotic analysis of Eq. \ref{eq:coupled} that, in the limit $\varepsilon_f\rightarrow0$, leads to writing the homogenized dynamics for the slow variables $(T,S)$ in terms of a stochastic differential equation. The presence of a two-way coupling between the fast and the slow variables requires a more general approach \cite{Engel2021} than the more common case of skew-symmetric systems, where the dynamics of the fast variables has an impact on the slow variables but not the other way around \cite{Kelly2011,Gottwald2013}. See \ref{multiscale} for a sketch of the asymptotic multiscale analysis of the system given in Eq. \ref{eq:coupled} and a derivation of the corresponding homogenized evolution equations for the slow variables.

\section{Chaotic transients} \label{sec:transients}
\begin{figure}[htbp]
    \centering
    \includegraphics[width=0.6\textwidth]{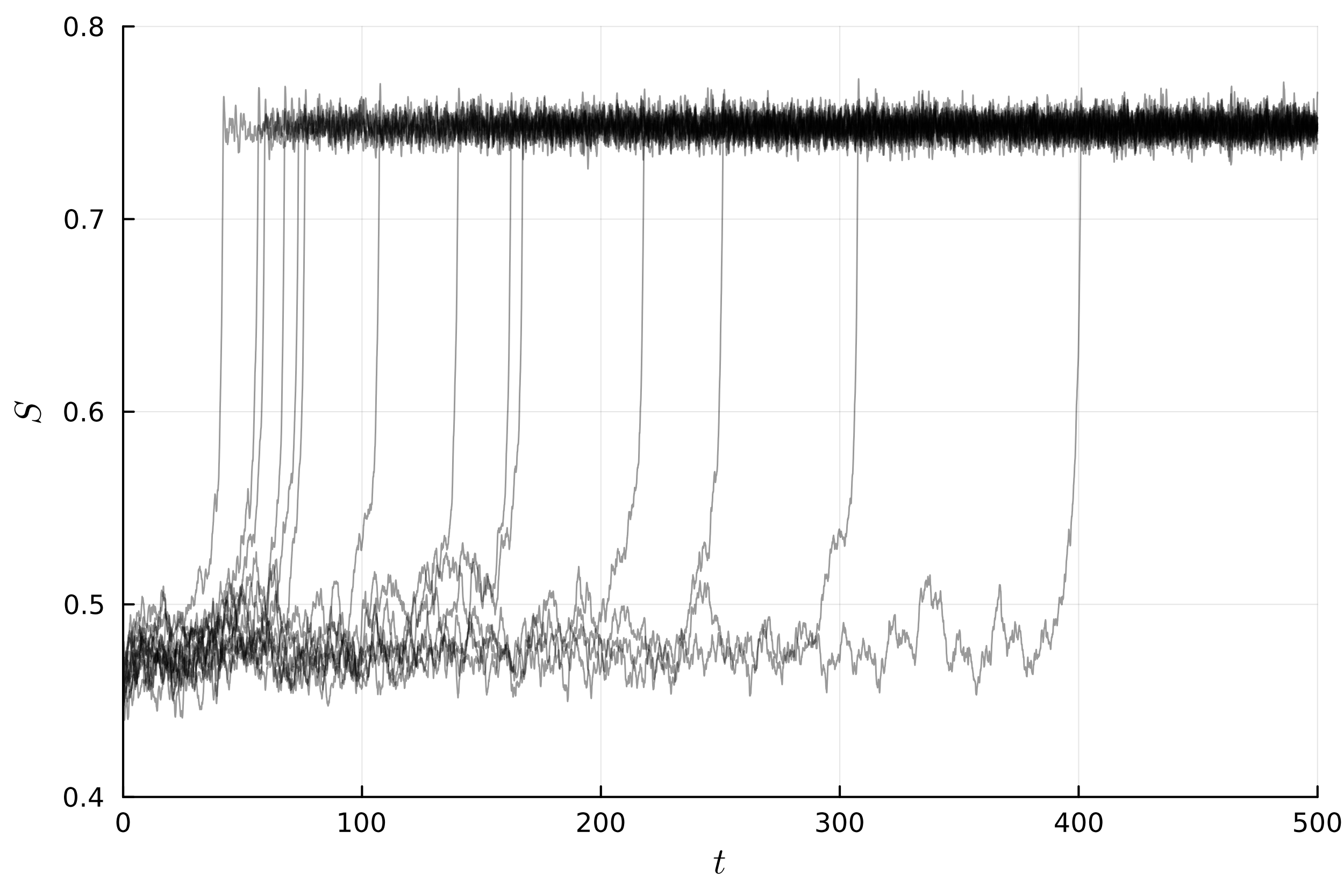}
    \caption{An example of long transients for $\sigma_0=0.932$. Initial conditions were sampled from the ``AMOC on'' attractor for $\sigma_0=0.926$ and were integrated forward after the parameter change. The lifetime of the transients spans one order of magnitude and appears unpredictable, but eventually all trajectories convergence to the only remaining attractor.}
    \label{fig:transient-example}
\end{figure}

We first turn our attention to the limits of predictability caused by the sensitivity of the lifetime of chaotic transients to the initial condition. As an example, we select a set of initial conditions by sampling the ``AMOC on'' state for $\sigma_0 = 0.926$ every $\Delta t=7.5$. As we will show explicitly below (cf. Fig. \ref{fig:coupled-bifurcation}), at this value of $\sigma_0$ the ``AMOC on'' state is globally stable. Then, we integrate trajectories from these initial conditions but with the freshwater parameter set to $\sigma_0 = 0.932$ (Fig. \ref{fig:transient-example}), i.e. slightly larger than for the bifurcation point $L_2$ of the uncoupled Stommel model, $\sigma(L_2)=0.9263$. We can assume that this value is also close to the bifurcation point of the coupled model because the coupling is weak (see Fig. \ref{fig:coupled-bifurcation}).

The initial conditions in Fig. \ref{fig:transient-example} remain in the vicinity of a ``ghost attractor'' reminiscent of the ``AMOC on'' state for a long time (up to 400 time units $\approx$ 40 000 years) before converging to the only remaining genuine attractor, the ``AMOC off'' state. In this section, we systematically explore the lifetime of such chaotic transients and show that they can be observed over a wide range of $\sigma_0$.

For an ensemble of initial conditions, the lifetime of chaotic transients, denoted as $\tau$, is expected to be exponentially distributed \citep{Yorke1979} with a mean lifetime $\langle \tau \rangle$, if $\tau$ is sufficiently large \citep{Grebogi1986}. Indeed, this is the case for a 100-member initial condition ensemble from which the trajectories in Fig. \ref{fig:transient-example} were drawn, with $\langle \tau \rangle \approx 200$ time units (20 000 years). Grebogi et al. \citep{Grebogi1986} then demonstrated that typically, the mean lifetime follows a power law as one approaches the critical value of the control parameter $\sigma_{0,\text{c}}$, such that we expect:
\begin{align}
\langle \tau \rangle \propto | \sigma_0 - \sigma_{0,\text{c}} | ^ {-\gamma} \label{eq:tau-powerlaw}
\end{align}
where $\gamma$ is the critical exponent.

Here, we characterize this region of transient chaos by evaluating the lifetimes of trajectories near the ``ghost attractor'' using an ensemble of initial conditions for different values of $\sigma_0$. The initial conditions are sampled over $t=200$ from the ``off'' or ``on'' state near the bifurcation points of the underlying Stommel model, and the results are not sensitive to the exact location of the sampling. Here and in the following, all sampling intervals are chosen to be at least several times larger than the Lyapunov timescale of the fast system ($\approx 0.002$ time units). Because $\langle \tau \rangle \to \infty$ as $\sigma_0 \to \sigma_{0,\text{c}}$, we need to choose a threshold time $T_\text{max}$ after which we deem a trajectory ``stable'', even though it may still be a very long chaotic transient. This means that $\sigma_{0,\text{c}}$ would be biased if determined by direct simulation, as it would depend on the choice of $T_\text{max}$. Therefore, we look to find a range of values for $\sigma_0$ with $\langle \tau \rangle$ sufficiently large for the values to be exponentially distributed, but smaller than $T_\text{max}$. Using Eq. \ref{eq:tau-powerlaw}, we then determine $\sigma_{0,\text{c}}$ and $\gamma$ within this region via a weighted least-squares fit, taking into account that the parameter $\langle \tau \rangle$ is distributed as $2n\hat{\langle \tau \rangle}/\chi^2(2n)$ \citep{Epstein1953}, where $\hat{\langle \tau \rangle}$ is equal to the sample mean of the lifetimes, $n$ is the number of samples, and $\chi^2(2n)$ is the chi-square distribution with $2n$ degrees of freedom.

\begin{figure}[htbp]
    \centering
    \includegraphics[width=\textwidth]{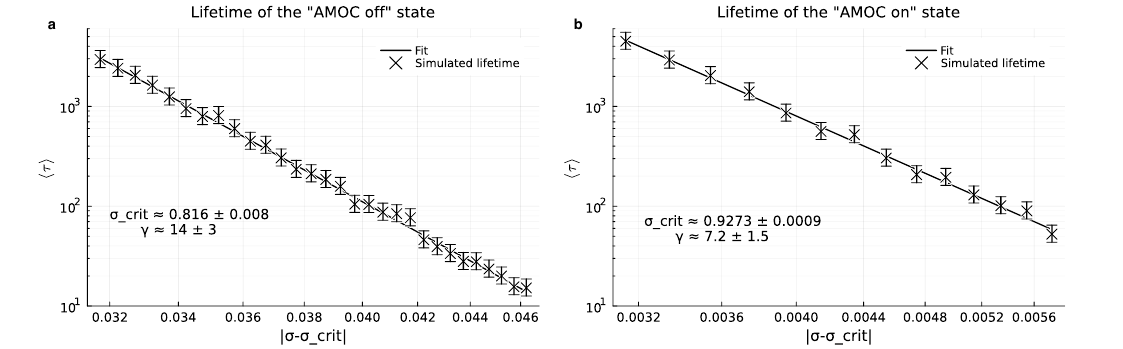}
    \caption{Numerically estimated mean lifetime $\langle \tau \rangle$ of chaotic transients, corresponding 95\% confidence intervals and power-law fit for different values of $\sigma_0$ for (a) the transient ``AMOC off'' state and (b) the transient ``AMOC on'' state outside the regime of bistability. Note that all axes are logarithmic and that the subplots use different $x$-axes.}
    \label{fig:lifetime-transient}
\end{figure}

Fig. \ref{fig:lifetime-transient} shows the lifetime of the transients tracking the ``ghost attractor'' as a function of the distance to the critical value $|\sigma_0 - \sigma_{0,\text{c}}|$. Here, we only evaluate the range of $\sigma_0$ in which the lifetimes of an ensemble of $n=100$ initial conditions approximately follow an exponential distribution, and in which $\tau < T_{max} = 10^4$ for all trajectories. Using the fitted critical value, there is a clear power-law relation (a straight line in the log--log plot) between the lifetime and $|\sigma_0 - \sigma_{0,\text{c}}|$ near both boundary crisis points. We obtain $\sigma_{0,\text{c}1} = 0.816 \pm 0.008$ and $\sigma_{0,\text{c}2} = 0.9273 \pm 0.0009$ (cf. Fig. \ref{fig:saddle-lifetime}b), where index ``1'' denotes the transition from the monostable ``AMOC on'' to the bistable regime, and index ``2'' denotes the transition from the bistable to the monostable ``AMOC off'' regime. These values are quite robust to the choice of the scaling law. For instance, fitting the non-power law scaling proposed by \citep{Grebogi1985},
%\begin{align}
%    \langle \tau \rangle \propto \text{exp}(k/| \sigma_0 - \sigma_{0,\text{c}} | ^ {1/2}), \label{eq:alternative-powerlaw}
%\end{align}
we obtain $\sigma_{0,\text{c}2} = 0.9251 \pm 0.0013$, which agrees well with the estimate from the fit to Eq. \ref{eq:tau-powerlaw}.

\begin{figure}[htbp]
    \centering
    \includegraphics[width=0.55\textwidth]{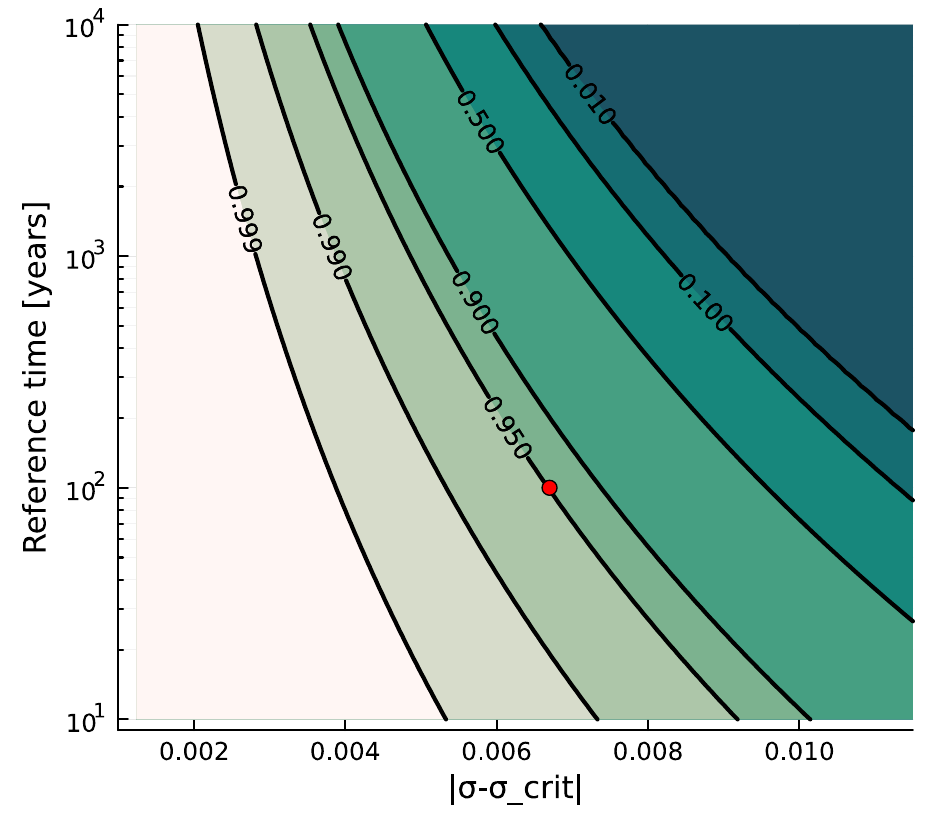}
    \caption{Probability of survival in the region of chaotic transients beyond the critical value $\sigma_{0,\text{c}2}\approx0.927$ as a function of the distance to the critical value $| \sigma_0 - \sigma_{0,\text{c}} |$ and the timescale of interest (reference time) during which tipping should be prevented. The red marker indicates $\sigma_0 \approx 0.934$ for which the transient tipping probability is limited to 5\% on a timescale of 100 years.}
    \label{fig:survival-probability}
\end{figure}

The critical exponents are $\gamma_1 = 14 \pm 3$ at the first and $\gamma_2 = 7.2 \pm 1.5$ at the second bifurcation. This implies that long transients can be observed over a wider range of parameter values at the first compared to the second bifurcation \citep{Ott2002}. For instance, the parameter range $|\sigma_0 - \sigma_{0,\text{c}}|$ for which $\langle \tau \rangle > 100$ is 0.04 at the first bifurcation and 0.005 at the second bifurcation. Nevertheless, both critical exponents are larger than those of many typical dynamical systems \citep{Ott2002}. Using the scaling law from Eq. \ref{eq:tau-powerlaw} and the quantile function of the exponential distribution, we can now calculate how far the ``safe operating space'' of the AMOC is extended beyond the genuinely bistable regime through the existence of long transients. In other words, up to which value of $\sigma_0$ will the desirable AMOC state (in our case, the ``on'' state) continue to act like a stable attractor, even though the bifurcation point has been crossed? The exponential distribution underlying the transient lifetime means that this safe operating space is inherently linked to a timescale of interest and an acceptable probability of survival on the desired AMOC state (or, equivalently, a tolerable probability of tipping to the undesired state). Fig. \ref{fig:survival-probability} shows this probability of survival on the ``on'' state as a function of $|\sigma_0 - \sigma_{0,\text{c}}|$ and the chosen timescale. For example, overshooting to $\sigma_0 \approx 0.934$ would yield a 5\% probability of obtaining a transient lifetime of less than 100 years (red marker in Fig. \ref{fig:survival-probability}). On the other hand, if the probability of a ``transient collapse'' should be limited to 0.1\% on the same timescale, the safe operating space is limited to no more than $\sigma_0 \approx 0.931$. In this way, if the control parameter is changed sufficiently rapidly to bring the system back into the bistable regime, a transition might successfully be avoided  despite ``overshooting'' the critical value (cf. \citep{Ritchie2021}).

\section{Chaotic saddle and fractal basin boundary} \label{sec:saddle}
We now focus on the limits to predictability in the bistable regime. The long-lived transients make it difficult to estimate its boundaries precisely, such that we rely on the two extrapolated critical values from Sec. \ref{sec:transients} to bound this regime ($\sigma_0 \in [0.82,0.927]$ in the following), within which the ``AMOC on'' and ``AMOC off'' states are expected to coexist as genuine attractors. As outlined above, in the bistable regime predictability of the asymptotic state is limited by the dimension $D_b$ of the fractal basin boundary. $D_b$ is in turn fundamentally linked to the properties of the chaotic saddle between the two attractors \citep{Hunt1996,Sweet2000,Bodai2020}. However, the chaotic saddle is typically a set with Lebesgue measure zero that has one unstable and one stable manifold (the basin boundary) \citep{Ott2002}, and therefore cannot be obtained by direct simulation. Therefore, we start with constructing a \textit{pseudo}-trajectory of the saddle (Sec. \ref{sec:edge-tracking}) which we use to determine its lifetime (Sec. \ref{sec:saddle-lifetime}) and Lyapunov exponents (Sec. \ref{sec:saddle-lyapunov}) before assessing the resulting basin boundary dimension (Sec. \ref{sec:basin-boundary}).

\subsection{Constructing the chaotic saddle: edge tracking algorithm} \label{sec:edge-tracking}
\begin{figure}[htbp]
    \centering
    \includegraphics[width=0.7\textwidth]{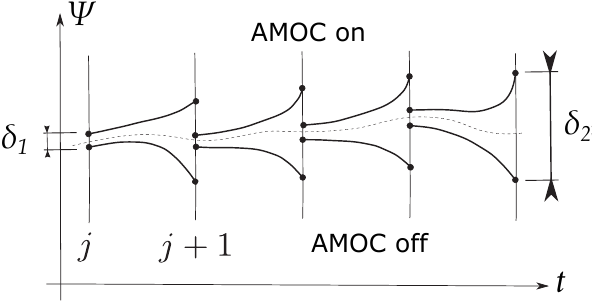}
    \caption{Schematic depiction of the edge tracking algorithm. Note that in this study, the distances $\delta_{1,2}$ are calculated over the entire phase space instead of only $\Psi$. Adapted from Lucarini and Bódai \citep{Lucarini2017}, released under a Creative Commons license (CC-BY 3.0, \url{https://creativecommons.org/licenses/by/3.0/}).}
    \label{fig:edge-tracking-alg}
\end{figure}

While there are different algorithms to construct a chaotic saddle approximately \citep{Lai2011}, we follow the strategy of \citep{Lucarini2017} by using the edge tracking algorithm originally proposed in \citep{Battelino1988} and later re-formulated by \citep{Skufca2006}. This allows us to obtain an arbitrarily long pseudo-trajectory that tracks the chaotic saddle of the conceptual atmosphere--ocean model. The edge tracking algorithm has previously been applied successfully in both high- and low-dimensional chaotic systems (e.g., \citep{Lucarini2017,Skufca2006,Cassak2007,Vollmer2009,Gelbrecht2021}). The algorithm is depicted schematically in Fig. \ref{fig:edge-tracking-alg} and described in detail in Sects. 4.2 and 5.1 of \citep{Lucarini2017}, such that we only summarize it briefly here.

We start from two initial conditions $\{\mathbf{u}_1^0, \mathbf{u}_2^0\}$ inside the two different basins of attraction. First, by applying the standard bisection method%by repeatedly replacing one state by $(\mathbf{u}_1^0+\mathbf{u}_2^0)/2$
, a pair of states $\{\mathbf{u}_1^\ast, \mathbf{u}_2^\ast\}$ with $|\mathbf{u}_1^\ast - \mathbf{u}_2^\ast| < \delta_1$ is obtained, in which $\mathbf{u}_1^\ast$ and $\mathbf{u}_2^\ast$ still belong to two different basins of attraction. By construction, these two states are both within a distance $\delta_1$ from the basin boundary. As a consequence, when the system is evolved from $\mathbf{u}_i(t_0) \equiv \mathbf{u}_i^\ast$, they are expected to diverge along the unstable direction of the basin boundary while tracking with the flow along its stable direction. Once $|\mathbf{u}_1(t) - \mathbf{u}_2(t)| \geq \delta_2$ for some $t>t_0$, the bisection is repeated such that the distance between the two ``shadowing trajectories'' $\mathbf{u}_1(t)$ and $\mathbf{u}_2(t)$ is again $< \delta_1$, and the new states are evolved further. This procedure can be repeated any number of times to obtain a pseudo-trajectory $\mathbf{u}_S(t) = (\mathbf{u}_1(t)+\mathbf{u}_2(t))/2$ that follows the basin boundary and eventually converges to track the chaotic saddle itself.

In the following, we set $\delta_2 = 0.004$ and $\delta_1 = 0.0025$, such that only one bisection step is needed per iteration. However, as in more complex models, it is not clear a priori how the norm $|\mathbf{u}_1(t) - \mathbf{u}_2(t)|$ should be defined, as the state vector $\mathbf{u}$ comprises different model components with possibly different magnitudes for mean and variability. Here, we define the norm by rescaling the atmospheric variables with a scaling factor $\xi < 1$ (we choose $\xi = 1/5000$). It can be viewed as a hyperparameter of our method and does not alter the dynamics of the dynamical system. This enables us to calculate the divergence of trajectories over the full phase space (instead of just the $T-S$ subspace), but ensures that the divergence is not dominated by the fast chaotic motion. We run the edge tracking algorithm for 2000 iterations, generating around 50 time units ($\approx 5000$ years) of pseudo-trajectory.

\begin{figure}[htbp]
    \centering
    \includegraphics[width=0.55\textwidth]{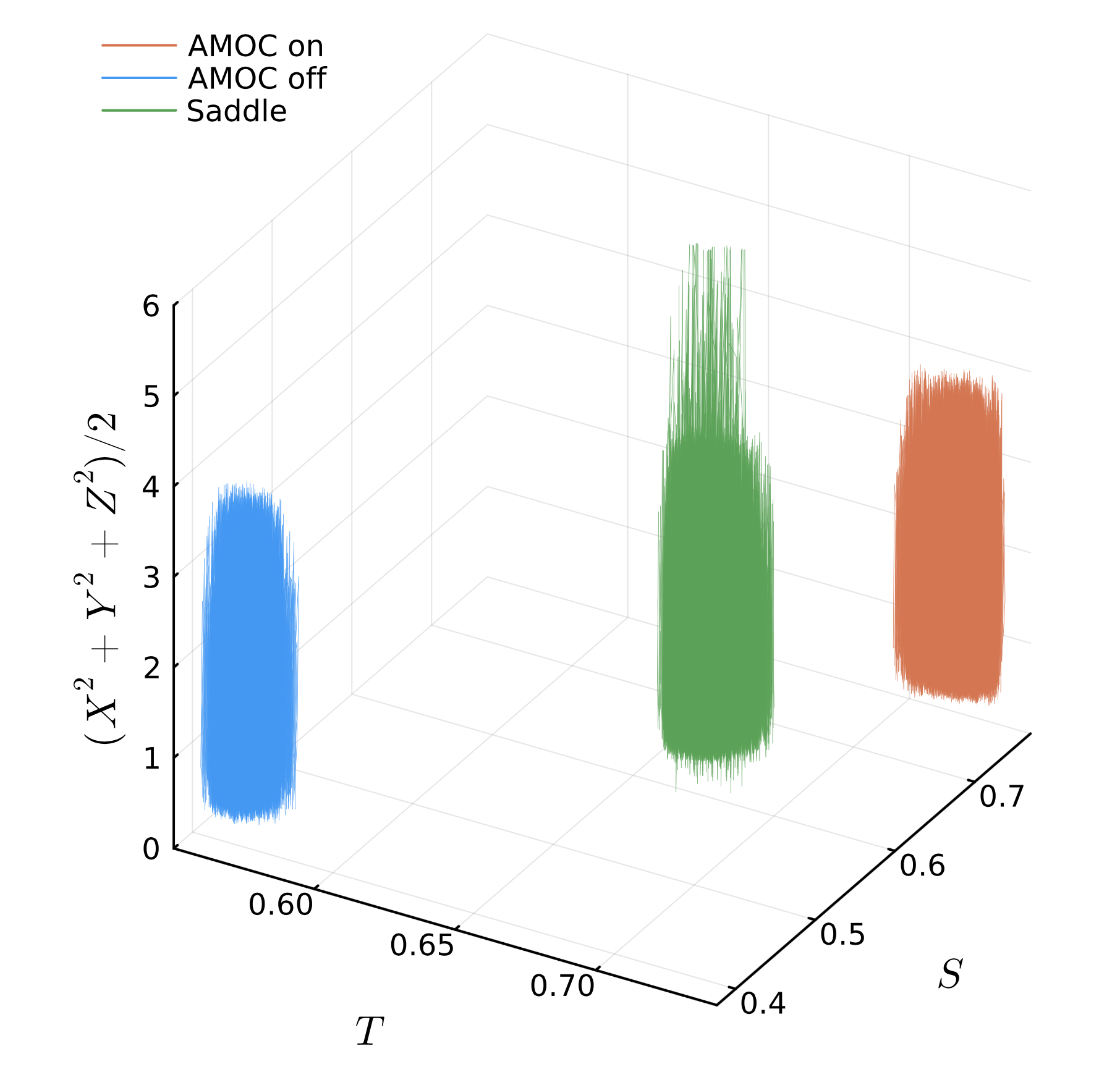}
    \caption{Three-dimensional projection of a pseudo-trajectory of the saddle (green) and trajectories tracking the attractors (blue, orange) for $\sigma_0 = 0.9$. Axis show the slow variables $T$ and $S$ and the atmospheric energy $E = (X^2 + Y^2 + Z^2)/2$.}
    \label{fig:saddle-sample}
\end{figure}

Fig. \ref{fig:saddle-sample} shows an example result of the edge tracking algorithm for $\sigma_0 = 0.9$. While the attractors and the saddle separate clearly in both $T$ and $S$ as expected from the phase space structure of the underlying Stommel model, the atmospheric motion differs little between the two attractors. However, close to the saddle the atmospheric energy $\frac{1}{2}(X^2 + Y^2 + Z^2)$ has longer tails, which might be due to the repelling along the unstable direction, and therefore a sign that the weak coupling does exert a certain influence on the atmosphere.

\begin{figure}[htbp]
    \centering
    \includegraphics[width=0.7\textwidth]{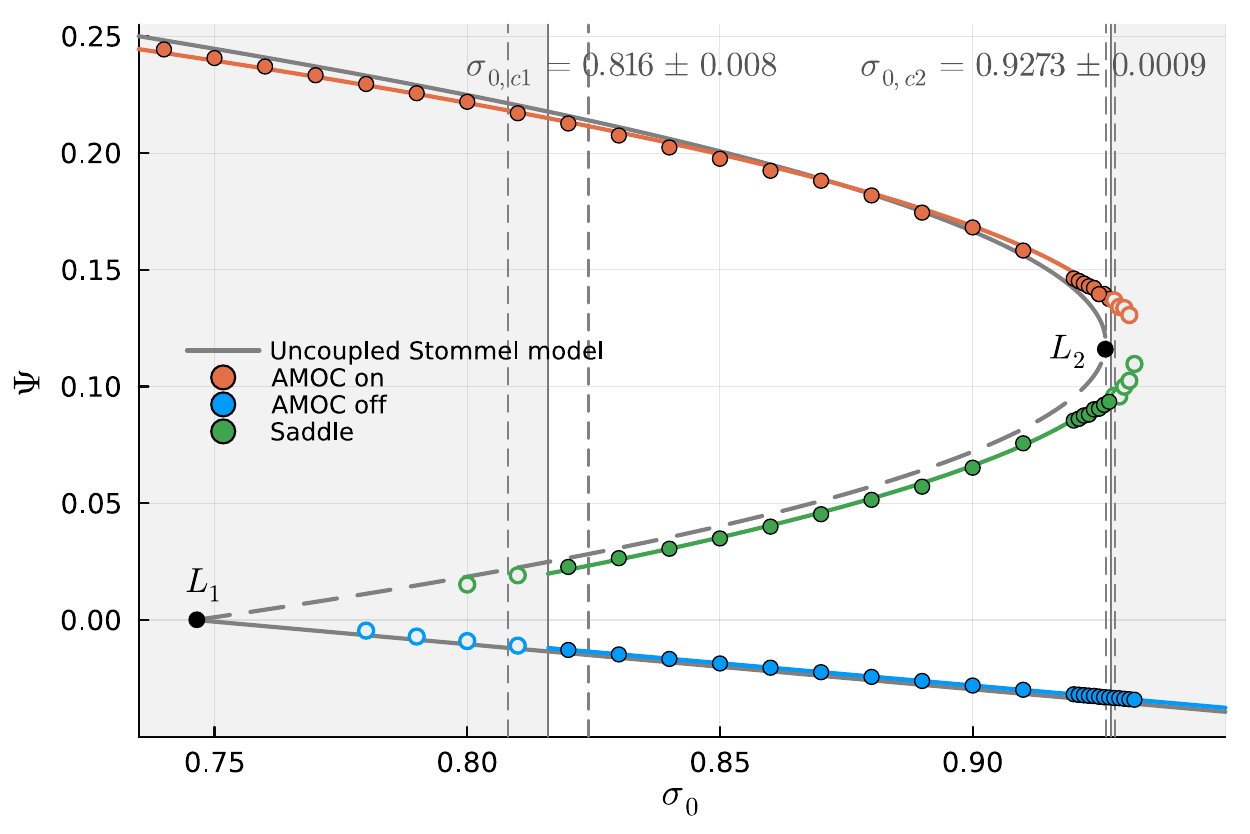}
    \caption{Bifurcation diagram for AMOC strength $\Psi$ of the coupled model. Each dot represents one (pseudo-)trajectory averaged over about 50 time units after a spin-up period. Points in the transient regime (gray background) as determined in Sec. \ref{sec:transients} are marked with unfilled circles. Vertical solid and dashed lines represent best estimates $\pm$ one standard deviation of the critical values $\sigma_{0,\text{c}i}$. The colored, solid lines represent square-root fits for the upper and the unstable branches and a linear fit for the lower branch. The bifurcation diagram of the uncoupled Stommel model (Eq. \ref{eq:stommel}) is shown in gray for comparison. Saddle-node bifurcation points for the Stommel model are labeled $L_1$ and $L_2$.}
    \label{fig:coupled-bifurcation}
\end{figure}

Applying the edge tracking algorithm for different values of $\sigma_0$, we can compute the unstable branch of the AMOC bifurcation diagram for the coupled atmosphere--ocean model (Fig. \ref{fig:coupled-bifurcation}). Over most of the parameter range, our bifurcation diagram follows that of the underlying uncoupled Stommel model very closely, as expected from weak atmosphere--ocean coupling combined with a large timescale separation. Outside the bistable parameter range, we can also perform edge tracking in a region of parameter space where very long-lived chaotic transients exist; in this case, the algorithm detects an edge state lying in between the long-lived chaotic solution and the actual unique asymptotic solution (unfilled circles in Fig. \ref{fig:coupled-bifurcation}). Note that in this regime, the approach is very similar to the one taken in turbulence \citep{Skufca2006}, where one finds the unstable solution between the laminar fixed point and the long-lived turbulent state. An interesting feature of the bifurcation diagram in Fig. \ref{fig:coupled-bifurcation} is that the stable and unstable branches do not meet at the critical values $\sigma_{0,\text{c}i}$. This might put into question the extrapolated critical values from Sec. \ref{sec:transients}, but we have verified the transient nature of some of the unfilled points on the ``AMOC on'' branch of Fig. \ref{fig:coupled-bifurcation} via direct simulation (e.g., Fig. \ref{fig:transient-example}). In these cases, the transient lifetime is longer than the length of the edge tracking trajectory ($t\approx50)$. Therefore, the discrepancy seems to arise from the particular multiscale nature of our system, which merits further investigation in future studies.

\subsection{Lifetime of the saddle} \label{sec:saddle-lifetime}
We now focus on two properties of the chaotic saddle, its lifetime and its spectrum of Lyapunov exponents (LEs) \citep{Ott2002}, for different values of the freshwater forcing $\sigma_0$ in the bistable regime. Both can be obtained by sampling a large number of initial conditions from the edge tracking trajectory obtained in Sec. \ref{sec:edge-tracking}. Since a state initialized on the trajectory $\mathbf{u}_S$ is very close to but not precisely on the actual saddle, it will remain in a phase space region $\mathcal{B}$ containing the saddle for some time before converging to one of the attractors.

Defining $\mathcal{B}$ as the bounding box of $\mathbf{u}_S$ after the initial spin-up, the number of remaining trajectories within $\mathcal{B}$ is expected to decay exponentially \citep{Ott2002}:
\[ N_S(t) = N_0\,e^{-t/ \langle \tau \rangle}, \]
where $N_0$ is the number of sampled initial conditions ($N_0=600$ in our numerical simulations), which are all within $\mathcal{B}$ by construction, and $\langle \tau \rangle$ is the mean lifetime. In practice, this exponential scaling is not expected to hold for small and large $t$ due to the non-uniform initial distribution in $\mathcal{B}$ and the finite size of the ensemble, respectively. However, we can obtain $\langle \tau \rangle$ via a least-squares fit of the slope of $\log N_S(t)/N_0$ against $t$ for intermediate values of $t$.

\begin{figure}[htbp]
    \centering
    \includegraphics[width=\textwidth]{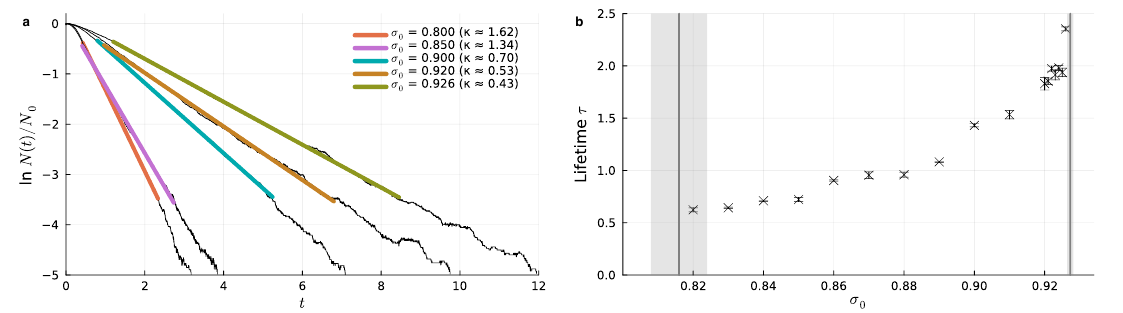}
    \caption{Lifetimes of the chaotic saddle. (a) Fraction of trajectories $N(t)/N_0$ within the saddle bounding box $\mathcal{B}$ as a function of time for different values of $\sigma_0$. Colored lines indicate linear fits in lin--log space, whose slope is $-\kappa$, in the range $N/N_0 \in [0.03, 0.7]$. (b) Lifetime $\tau = 1/\kappa$ obtained from the fits in panel a for all simulated values of $\sigma_0$. Error bars reflect the uncertainty in $\tau$ with respect to reasonable changes in the fitting range in panel a. The vertical lines and shaded areas indicate estimates of the critical values $\sigma_{0,c}$ plus and minus one standard deviation, corresponding to the vertical solid and dashed lines in Fig. \ref{fig:coupled-bifurcation}.
    }
    \label{fig:saddle-lifetime}
\end{figure}

Fig. \ref{fig:saddle-lifetime}a shows $N_S(t)/N_0$ for several values of the freshwater forcing $\sigma_0$ in the bistable regime of the Stommel model. As expected, the number of remaining trajectories in $\mathcal{B}$ shows a clear exponential decay. The lifetime is of $\mathcal{O}(1)$ (about one century in physical units) but depends strongly on $\sigma_0$: approaching the saddle-node bifurcation of the underlying Stommel model, trajectories spend more time in the vicinity of the saddle. While the number of sampled $\sigma_0$ is too small to conclusively establish a functional relation $\langle \tau \rangle(\sigma_0)$, the lifetime increases clearly non-linearly as the critical freshwater forcing parameter $\sigma_{0,c2}$ is approached (Fig. \ref{fig:saddle-lifetime}b).

\subsection{Lyapunov spectrum of the saddle} \label{sec:saddle-lyapunov}
Having explored the global instability of the coupled model characterized by the lifetime of the saddle, we now turn towards the local instability due to its chaotic dynamics. To this end, we compute the full spectrum of Lyapunov exponents of the saddle using the standard procedure of successive Gram-Schmidt orthonormalization \citep{Benettin1980} as implemented by \citep{Datseris2018}. The Lyapunov spectrum of the saddle is approximated by choosing 2000 initial conditions on $\mathbf{u}_S$ with the longest lifetimes in $\mathcal{B}$ and then averaging over the Lyapunov spectra of a subset of these individual trajectories while they remain in $\mathcal{B}$. For consistency, we tested the same method for the attractors, where averaging over 2000 finite-time LEs calculated over a similar duration ($t = 7.5$) yielded a comparable result to letting the algorithm converge along one long trajectory.

\begin{table}[htbp]
    \centering
    \begin{tabular}{l|c|c|c}
         & ``AMOC on'' & ``AMOC off'' & Saddle \\
         \hline
        $\lambda_1$ & $535 \pm 2$ & $531 \pm 2$ & $538 \pm 2$ \\
        $\lambda_2$ & $-0.04 \pm 0.02$ & $-0.015 \pm 0.017$ & $0.13 \pm 0.07$ \\
        $\lambda_3$ & $-0.67 \pm 0.02$ & $-2.65 \pm 0.03$ & $-0.56 \pm 0.14$ \\
        $\lambda_4$ & $-8.11 \pm 0.13$ & $-2.91 \pm 0.04$ & $-6.0 \pm 0.3$ \\
        $\lambda_5$ & $-1230 \pm 2$ & $-1155 \pm 2$ & $-1182 \pm 4$ \\
    \end{tabular}
    \caption{Lyapunov exponents for the two attractors (labeled as in Fig. \ref{fig:saddle-sample}) and the chaotic saddle for $\sigma_0=0.9$. Uncertainty estimates are 95\% confidence intervals generated from bootstrap resampling of 100 trajectories each from the 2000-member ensemble.}
    \label{tab:lyapunov-spectrum}
\end{table}

The Lyapunov spectra $\{\lambda_i\}, i=1,...,5$ for the saddle and each of the two attractors are given in Tab. \ref{tab:lyapunov-spectrum} for $\sigma_0 = 0.9$. For the attractors, we expect one positive LE characterizing the chaotic instability on the attracting set, a zero LE associated with the motion along the attractor, and otherwise negative LEs encapsulating the convergence of initial conditions to the attractor. Indeed, we find a large, positive maximum Lyapunov exponent (MLE) $\lambda_1$, while $\lambda_2$ is very close to zero and all remaining LEs are negative. Similarly, the saddle possesses one positive LE ($\lambda_1$), one LE close to zero ($\lambda_2$) and three negative LEs. Note that obtaining a good numerical estimate for the vanishing LE of a saddle is more difficult than for an attractor. This is because the trajectories considered for the evaluation of the LEs have a finite lifetime, as they unavoidably end up veering towards one of the competing attractors, realizing a positive stretching along the direction of the flow. We have verified (Fig. S1 in the Supplementary Information [SI]) that the estimate of $\lambda_2$ converges towards zero as we consider longer-lived trajectories for our estimates, so that the stretching along the flow is minimized (being zero if we could generate a trajectory living exactly on the saddle).

%Note also that the uncertainties in Tab. \ref{tab:lyapunov-spectrum} reflect the aleatoric uncertainties due to sampling different sets of initial conditions, but do not take into account any systematic uncertainties that arise from the finite sampling time or the large timescale separation in our system. Hence, we are confident that $\lambda_2$ corresponds to the vanishing LE in all cases.

The MLE of our model can be attributed to the chaotic motion induced by the atmosphere via a simple scaling test. First, we note that the magnitude of the maximum and minimum LEs ($|\lambda_1|$ and $|\lambda_5|$) is large compared to the other LEs. Furthermore, the values of $\lambda_1$ and $\lambda_5$, respectively, are similar for all three invariant sets. This is consistent with the assumption that the atmospheric and oceanic components of our model are weakly coupled, and we may associate $\lambda_1$ and $\lambda_5$ with the maximal and minimal LE of the uncoupled L84 model, rescaled with the atmospheric timescale $1/\varepsilon_f$. This can be confirmed by computing the Lyapunov spectrum of the attractors for different values of $\varepsilon_f$ (Fig. S2 in the SI). We find that $\lambda_1$ and $\lambda_5$ scale in very good agreement with $\lambda_{i,\text{L84}}/\varepsilon_f$, where $\lambda_{i,\text{L84}}$ is the corresponding LE of the uncoupled L84 model from Eq. \ref{eq:L84}, while the second to fourth LEs are mostly independent of the timescale separation.

The results shown here exemplarily for $\sigma_0=0.9$ are qualitatively representative for all values of $\sigma_0$ which we have tested within the bistable regime, and there is no clear dependence of $\lambda_1$ or $\lambda_2$ on the freshwater flux. The MLE $\lambda_1$ is very similar across different $\sigma_0$ (mean: 536 $\pm$ 4) and the second LE is very small ($\lambda_2 < 0.3 \ll \lambda_1$) for all $\sigma_0$.

\subsection{Dimension of the fractal basin boundary} \label{sec:basin-boundary}
Given the spectrum of Lyapunov exponents and the lifetime of the saddle, we can use the dimension formula conjectured by \citep{Hunt1996,Sweet2000} to relate these two invariants to the information dimension of the stable manifold of the saddle, i.e., the basin boundary. However, provided that the escape rate from the saddle $\kappa = 1/\tau$ is much smaller than the MLE $\lambda_1$, Bódai and Lucarini \citep{Bodai2020} showed that the formula for the dimension of the basin boundary $D_b$ can be simplified to (see their Eq. 25)
\begin{align}
    D_b = D - \frac{\kappa}{\lambda_1} \,. \label{eq:boundary-dim}
\end{align}
This relation can be viewed as a generalization of the formula proposed earlier by Hsu et al. \citep{Hsu1988} to the case of rough basin boundaries \citep{Bodai2020}. Note that, strictly speaking, Eq. \ref{eq:boundary-dim} gives the information dimension and not the box-counting dimension of the basin boundary which will be needed later for assessing the question of predictability, but the two are expected to yield very similar values here \citep{Hunt1996}.

Since the Lyapunov spectra and lifetimes obtained above clearly satisfy $\kappa < \lambda_1$, we can apply Eq. \ref{eq:boundary-dim} to obtain a theoretical value for the basin boundary dimension. The resulting dimension $D_b$ is between $4.997$ and $4.999$ for different values of $\sigma_0$, with slightly increasing values as the critical values $\sigma_{0,\text{c}2}$ is approached (Fig. S3 in the SI). This means that the basin boundary dimension is extremely close to (within 0.1\%) but strictly smaller than the phase space dimension $D=5$. Following Eq. \ref{eq:boundary-dim}, this can be attributed to the escape rate being small compared to the MLE.

\begin{figure}[htbp]
    \centering
    \includegraphics[width=0.7\textwidth]{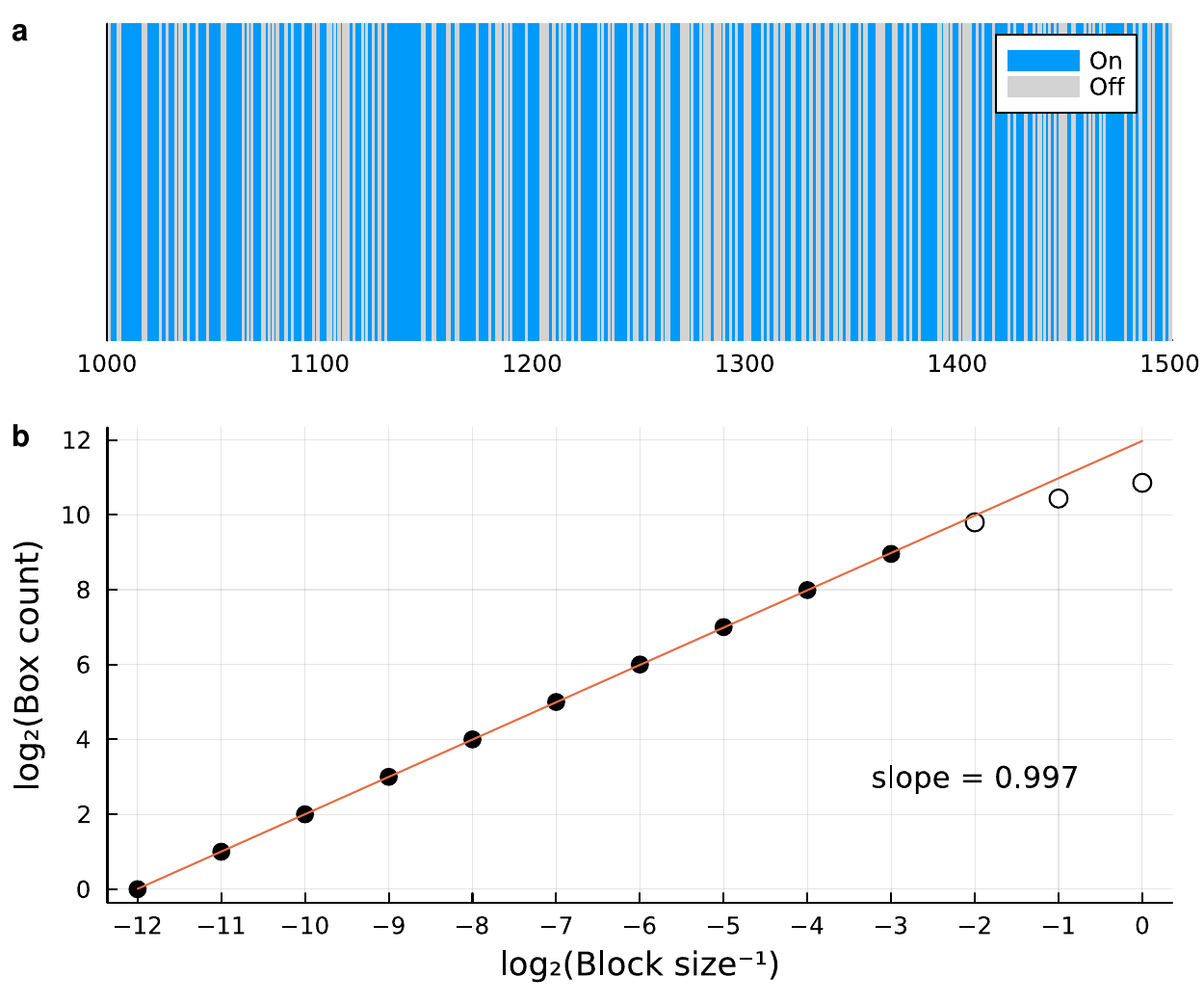}
    \caption{Fractal basin boundary along a sample cross-boundary section for $\sigma_0 = 0.9$: (a) Outcome (final AMOC state, color-coded) of the forward integration of 512 evenly spaced initial conditions along a one-dimensional phase space section intersecting with the fractal basin boundary (see Fig. S4 in the SI for a visualization of all outcomes), (b) Computation of the box-counting dimension from $2^{13}$ initial conditions from the same boundary section. Here, the ``pruning'' parameter is $N_p = 3$, i.e., the three rightmost points (unfilled markers) are not taken into account for the linear fit.}
    \label{fig:boundary-intersection-boxcount}
\end{figure}

To verify these results numerically, we also compute the dimension of the basin boundary directly using the standard box-counting algorithm. We follow the methodology of \citep{Lucarini2017} and \citep{Bodai2020} in sampling evenly spaced initial conditions along a line that intersects with the basin boundary, and determine the attractor to which each initial condition eventually converges. In Fig. \ref{fig:boundary-intersection-boxcount}a, we show a sample plot of the outcomes of $2^{13}$ initial conditions, where the endpoints of the sampling interval are within $\delta_1 = 10^{-3}$ of the basin boundary (using the definition of the norm from Sec. \ref{sec:edge-tracking}), but away from the saddle. Repeating this procedure in different regions of phase space and then for each value of $\sigma_0$ considered previously, we find that the standard box-counting dimension ($0 < D_\text{box} \leq 1$) as shown exemplarily in Fig. \ref{fig:boundary-intersection-boxcount}b is very close to one (between 0.988 and 0.999) for all $\sigma_0$. Both methods therefore confirm, regardless of whether the information or box-counting dimension is used, that a large scale separation between $\kappa$ and $\lambda_1$ yields a basin boundary with almost full phase space dimension.

The basin boundary dimension $D_b$ and the uncertainty exponent $\alpha$ that determines the \textit{final state sensitivity} as given by Eq. \ref{eq:predictability} are linked by the simple relation $\alpha = D - D_b$ \citep{McDonald1985}. With the values of $D_b$ predicted from the saddle properties (Fig. S3 in the SI), $\alpha$ varies between 0.003 and about 0.001 close to the critical value $\sigma_{0,\text{c}2}$. From the viewpoint of Eq. \ref{eq:predictability}, this means that an improvement in accuracy of the initial condition has a negligible effect on the uncertainty of the final climate state: if the accuracy of the initial condition is doubled, the phase space volume in which the outcome is uncertain decreases by about $1 - \left(\frac{1}{2}\right)^\alpha \approx 0.1$--$0.2\%$. Even if slightly larger values of $\alpha$ obtained from the box-counting algorithm are used, which are often on the order of $10^{-2}$, predictability of the final state only increases by 1--2\% for a doubling in the accuracy of the initial condition. Note that, formally, this vanishing predictability refers to the full five-dimensional system, but the two competing attractors separate mostly along the oceanic variables $(T,S)$ (cf. Fig. \ref{fig:saddle-sample}). Hence, we can interpret our results as vanishing predictability of the AMOC state. In summary, the final AMOC state is essentially unpredictable in an extended region of phase space close to the basin boundary, to first order irrespective of the value of $\sigma_0$.

\section{Discussion} \label{sec:discussion}
Using a conceptual climate model that comprises a chaotic atmosphere and a bistable AMOC, we have investigated two fundamental limits to predictability of the asymptotic AMOC state: a fractal basin boundary in the bistable AMOC regime and chaotic transients in the monostable regime. In the bistable regime, we approximated the chaotic saddle (or \textit{Melancholia state}) between the two competing AMOC attractors using the edge tracking algorithm and found a large timescale separation between the fast chaotic motion on the saddle and the slow escape rate from the saddle. This timescale separation implies a fractal basin boundary with close to full phase space dimension, which we verified by explicitly computing the box-counting dimension of the boundary. 
In the monostable regime, chaotic transients with exponentially distributed lifetimes arise close to the bifurcation points over a relatively wide range of the freshwater parameter $\sigma_0$.

This complex behavior complicates the assessment of a system's resilience in two distinct ways. In dynamical systems theory \citep{Krakovska2023}, one common definition of resilience relies on measuring the minimal ``kick'' perturbation that causes the system to transition into a competing (undesired) attractor \citep{Holling1973}. A fractal basin boundary, as quantified by Eq. \ref{eq:predictability}, thus implies a complete loss of resilience in an extended region in state space, where an arbitrarily small perturbation may cause a critical transition. In parameter space, resilience is often defined via the distance to critical points delimiting a ``safe operating space'' \citep{Rockstrom2009,Krakovska2023}. The presence of long chaotic transients on ``ghost attractors'' renders it essentially impossible to determine the exact position of such bifurcation points on finite timescales by observing or simulating only a single time series. Hence, defining a safe operating space requires a probabilistic definition due to the exponential distribution of transient lifetimes, and predictability depends crucially on the timescale of interest.

The two phenomena described here have recently been framed as \textit{fractality-induced tipping} and \textit{transient-reduced tipping}, respectively, by Kaszás et al. \citep{Kaszas2019}. To our knowledge, this is the first study in which both phenomena have been systematically explored for a deterministic but chaotic (conceptual) climate model. In contrast to \citep{Kaszas2019}, we did not focus on individual \textit{tipping probabilities}, but rather on a global characterization of phase space through the uncertainty exponent $\alpha$ (in the bistable regime) and the transient lifetime $\langle \tau \rangle$ (in the monostable regime). In the bistable regime, this can be achieved with a much smaller computational cost than the local phase space sampling approach taken by \citep{Kaszas2019}. In fact, calculating $\alpha$ via the saddle properties and Eq. \ref{eq:boundary-dim} also appears to yield more accurate results than repeatedly sampling the box-counting dimension of the basin boundary. While both global and local approaches would ideally complement each other, the construction of the saddle and computation of its properties might be computationally feasible in GCMs \citep{Lucarini2017}, while a sufficiently large ensemble might not.

To our knowledge, this is the first time that the edge tracking algorithm has been applied to find the chaotic saddle between two competing AMOC states. While it is not guaranteed that our results from a conceptual climate model will generalize to (much) higher-dimensional atmosphere--ocean models, we note several similarities with the results of Lucarini \& Bódai \citep{Lucarini2017}, who used an intermediate-complexity climate model with $\mathcal{O}(10^4)$ degrees of freedom to study a different tipping point. From a technical standpoint, the edge tracking algorithm worked robustly, independently of the exact nature of the positive feedback and regardless of whether internal variability of the target variable was larger (as in our case) or smaller than $\delta_1$. This makes us confident that the edge tracking algorithm is a suitable method even for models with many more degrees of freedom in which bistability of the AMOC has been identified (e.g., \citep{Hawkins2011,Lohmann2021,Willeit2022,Mehling2023}), even to those featuring large internal variability \citep{Mehling2023}. Indeed, we have recently implemented the edge tracking algorithm successfully in a coupled atmosphere--ocean GCM of intermediate complexity to investigate a Melancholia state separating the two competing AMOC states \cite{boerner2024}. While other methods such as numerical continuation have previously been applied to find unstable AMOC equilibria \citep{Dijkstra2005}, the edge tracking method only requires (sufficiently long) forward integration, which makes it easy to apply and very suitable for tipping elements in explicit models like typical GCMs.

A perhaps surprising similarity to \citep{Lucarini2017} is that the basin boundary dimension is similarly close to full dimension despite the very different models and feedbacks. We believe that this is indeed the case for a rather wide class of models that feature (explicit or emergent) timescale separation, with the local instability governed by the fast component (atmosphere/weather) and the global instability governed by the slow component (ocean/climate). Thus, it seems plausible that sensitive and seemingly random dependence on the initial condition of the final AMOC state \citep{Lohmann2021,Cini2023} or of the transient lifetime of a weak AMOC state \citep{Romanou2023} may be linked to the presence of a high-dimensional chaotic saddle and a fractal basin boundary.

So far, we have only investigated the autonomous case, which is approached when a parameter (here, the freshwater flux $\sigma_0$) is varied very slowly compared to the internal timescales of the system. This assumption is, however, invalid for the current anthropogenic warming and associated changes in the water cycle. Therefore, rate-dependent AMOC tipping could be observed, but the inertia of the system can also give rise to ``safe overshoots'' \citep{Ritchie2021} beyond the effect of transient phenomena discussed here. It remains an intriguing open question how the barriers to final state predictability discussed here would be altered in a non-autonomous setting and how they would depend on the forcing rate.

\section{Conclusions}\label{sec:conclusions}
Our results show that predictability of the asymptotic AMOC state is limited by a fractal basin boundary with almost full phase space dimension as well as long chaotic transients, which both arise from the chaotic multiscale nature of the coupled atmosphere--ocean system. We derived these findings from a conceptual climate model that consisted of four main ingredients for this behavior -- bistability, chaotic motion, timescale separation and weak coupling --, such that our conclusions should generalize to more complex models with similar properties and even other components of the Earth system. This has practical implications for the boundaries of a safe operating space, which become intrinsically fuzzy both in a spatial and temporal sense. Starting from a state in the ``gray zone'' of phase space near a fractal basin boundary or near the ``ghost attractor'' in the parameter region of chaotic transients, the final outcome depends sensitively on the initial condition, which may give rise to non-monotonic and seemingly counter-intuitive outcomes of an initial condition ensemble -- interpretable, however, via dynamical systems theory.
%\end{linenumbers}

\section*{Declaration of competing interest}
The authors declare that they have no known competing financial interests or personal relationships that could have appeared to influence the work reported in this paper.

%\section*{Code and data availability}
%Source code will be made available in a public repository (\url{https://doi.org/10.5281/zenodo.10370900}) at latest upon acceptance and can be obtained via request to the corresponding author until then. Code and data to reproduce the figures can be obtained on request from the corresponding author.

\section*{Acknowledgements}
This project has received funding from the European Union’s Horizon 2020 research and innovation programme under the Marie Skłodowska-Curie grant agreement No 956170 (CriticalEarth) and under grant agreement No 820970 (TiPES). The authors would like to thank Tam\'as B\'odai, Georg Gottwald, Calvin Nesbitt and Raphael R\"omer for discussions on different parts of this work. The authors also wish to thank two anonymous reviewers for their constructive comments which have improved the quality of the paper.

\appendix

\section{Parameters of the coupled L84-Stommel model} \label{appendix:model-parameters}
\begin{tabular}{c|l|p{.6\textwidth}}
     Parameter & Value & Physical meaning \\
     \hline
     $a$ & 0.25 & – \\
     $b$ & 4 & – \\
     $F_0$ & 8 & Atmospheric meridional temperature gradient \\
     $G_0$ & 1 & Atmospheric zonal temperature gradient \\
     \hline
     $\varepsilon_a$ & 0.34 & Ratio between temperature restoring timescale and diffusive timescale of the ocean \\
     $\mu$ & 7.5 & Ratio between advective and diffusive ocean timescales \\
     $\theta_0$ & 1 & Sea surface temperature gradient \\
     $\sigma_0$ & is varied & Sea surface freshwater flux \\
     \hline
     $\theta_1$ & 0.0195 & Sea surface temperature gradient perturbation \\
     $\sigma_1$ & 0.00934 & Sea surface freshwater flux perturbation \\
     $\bar{x}$ & 1.0147 & Time-mean of $x$ of the uncoupled L84 \\
     $\bar{\Delta}$ & 1.7463 & Time-mean of $\Delta$ of the uncoupled L84 \\
     $F_1$ & 0.1 & Atmospheric meridional temperature gradient perturbation \\
     $\varepsilon_f$ & $3 \cdot 10^{-4}$ & Ratio between atmospheric and oceanic timescale
\end{tabular}

\section{Multiscale model reduction}\label{multiscale}
As discussed in Section \ref{sec:model}, $\varepsilon_f$ is a small parameter that controls the timescale separation between the dynamics of the atmospheric and oceanic component of the system. It is possible to cast Eqs. \ref{eq:coupled} in the following form,
\begin{align} \label{eq:multi}
\begin{split}
\dot{\mathbf{x}} &= \frac{1}{\epsilon^2}g(\mathbf{x},\mathbf{X}) = \frac{1}{\epsilon^2}g_1(\mathbf{x})+\frac{1}{\epsilon^2}g_2(\mathbf{X})\\
\dot{\mathbf{X}} &= a(\mathbf{x},\mathbf{X})+\frac{1}{\epsilon}b(\mathbf{x},\mathbf{X}) = a(\mathbf{X})+\frac{1}{\epsilon}b(\mathbf{x}) \,,
%\varepsilon_f \dot{y} &= x y - b x z - (y - G_0) \\
%\varepsilon_f \dot{z} &= b x y + x z - z \\
%\dot{T} &= -\frac{1}{\varepsilon_a}(T - T_\text{surf}) - T - \mu |S-T| T \\
%\dot{S} &= S_\text{surf} - S - \mu |S-T| S
\end{split}
\end{align}
with $\epsilon^2=\varepsilon_f$, where $\mathbf{x}=(x,y,z)$  and $\mathbf{X}=(T,S)$ are vectors comprising the fast and slow variables, respectively. Because of the non-vanishing term $g_2$, the system is not skew-symmetric. Hence, one cannot straightforwardly apply the homogenization theory presented in \cite{Kelly2011,Gottwald2013} to derive the effective stochastic differential equation (SDE) in It\^o convention,
\begin{equation}\label{eq:homo}
    \dd\mathbf{X}=\tilde{a}(\mathbf{X})\dd t + \sigma(\mathbf{X})\dd\mathbf{W} \,,
\end{equation}
describing the properties of the slow variables $X$ in the $\epsilon\rightarrow0$ limit. Here $\tilde{a}(\mathbf{X}) \in\mathbb{R}^2$ is the drift term, $\sigma(\mathbf{X})\in\mathbb{R}^{2\times p}$ is the matrix describing the noise law, and $\dd\mathbf{W} \in\mathbb{R}^2$ is a vector whose components are the increments of $2$ independent Wiener processes.

Instead, one needs to resort to a more general theory \cite{Engel2021} that is able to deal with the conceptually more challenging case of two-way coupled systems for rather general functions $g(\mathbf{x},\mathbf{X})$, $a(\mathbf{x},\mathbf{X})$, and $b(\mathbf{x},\mathbf{X})$ in Eq. \ref{eq:multi} above. We derive the following explicit expressions for the drift term and the noise law in Eq. \ref{eq:homo}:
\begin{equation}\label{eq:para}
\tilde{a}=a(\mathbf{X}) \ , \quad\quad \sigma(\mathbf{X})\sigma(\mathbf{X})^T=\int_0^\infty dt \int d\mu_\mathbf{X}(\mathbf{x})\left(b(\mathbf{x})\otimes b(\mathbf{x}(t))+b(\mathbf{x}(t))\otimes b(\mathbf{x}) \right) \,,
\end{equation}
where $\otimes$ indicates the tensor product, $\mu_{\mathbf{X}}(\mathbf{x})$ is the invariant measure of the system $\dot{\mathbf{x}} =g_1(\mathbf{x})+g_2(\mathbf{X})$, where $\mathbf{X}$ is taken as a fixed parameter, and $\mathbf{x}(t)$ is the evolution at time $t$ of the initial condition $\mathbf{x}(t=0)=\mathbf{x}$. Hence, the correlation matrix of the noise is built starting from the properties of correlation of the fast system where the slow variables $\mathbf{X}$ are \textit{frozen in}. While the general homogenization formulas presented in \cite{Engel2021} are more complex, the rather simple result given in Eq. \ref{eq:para} comes from the fact that in our case the dependence on the fast and slow variables factors out in the functions $g$, $a$, and $b$. Indeed, one obtains the same result one would derive by naively using the theory developed for skew-symmetric systems, under the heuristic assumption that the slow variables can be treated as external parameters for the stochastic process associated with the fast variables.

We also wish to point out the fundamental difference between the original system given in Eq. \ref{eq:coupled} and its homogenized version for the slow variables given here in Eq. \ref{eq:homo}. In the parametric range where multistability is found for \eqref{eq:coupled}, it is by definition impossible for a  trajectory initialized in the basin of attraction of one state to visit the competing state. By contrast, the stochastic differential equation given in Eq. \ref{eq:homo} allows for (typically rare, when far from the bifurcation points) noise-induced transitions between the competing states of the slow variables (see discussion in \cite{Lucarini2019,Lucarini2020,Galfi2021}).

\bibliographystyle{elsarticle-num}
\urlstyle{same}
\bibliography{plasim-lsg-amoc,newbibs}

\end{document}

\begin{frontmatter}
\abstracttitle{List of Figures}


\title{Supplementary material for:\\ Limits to predictability of the asymptotic state of the Atlantic Meridional Overturning Circulation in a conceptual climate model}

\author[1]{Oliver Mehling}
\author[2,3]{Reyk Börner}
\author[2,3,4]{Valerio Lucarini}

\affiliation[1]{organization={DIATI – Department of Environment, Land and Infrastructure Engineering, Politecnico di Torino},
             addressline={Corso Duca degli Abruzzi 24},
             postcode={10129},
             city={Turin},
             country={Italy}}
\affiliation[2]{organization={Department of Mathematics and Statistics},
             addressline={University of Reading},
             city={Reading},
             postcode={RG6 6AX},
             country={United Kingdom}}
\affiliation[3]{organization={Centre for the Mathematics of Planet Earth},
             addressline={University of Reading},
             city={Reading},
             postcode={RG6 6AX},
             country={United Kingdom}}
\affiliation[4]{organization={School of Mathematical and Computational Sciences},
             addressline={University of Leicester},
             city={Leicester},
             postcode={LE1 7RH},
             country={United Kingdom}}

\begin{abstract}
    \renewcommand{\listfigurename}{\vspace{-2\baselineskip}}

    {%
    \let\oldnumberline\numberline%
    \renewcommand{\numberline}{\figurename~\oldnumberline}%
    \listoffigures%
    }
\end{abstract}

\end{frontmatter}



%\clearpage

%\begin{figure}[htbp]
%    \centering
%    \includegraphics[width=1.0\textwidth]{figs/Lyapunov_spectrum_histogram_09.pdf}
%    \caption[Distributions of finite-time Lyapunov exponents $\lambda_{1..5}$ for $\sigma_0=0.9$]{Lyapunov exponents $\lambda_i$ for $\sigma_0=0.9$ of the two attractors and the saddle. The distributions are approximated via a kernel density estimate from the histograms of the individual Lyapunov spectra along each of the $\approx$100 short trajectories. Vertical lines show the means over all these trajectories.}
%    \label{sfig:hist-lyapunov-spectrum}
%\end{figure}

\clearpage

\begin{figure}[htbp]
    \centering
    \includegraphics[width=0.8\textwidth]{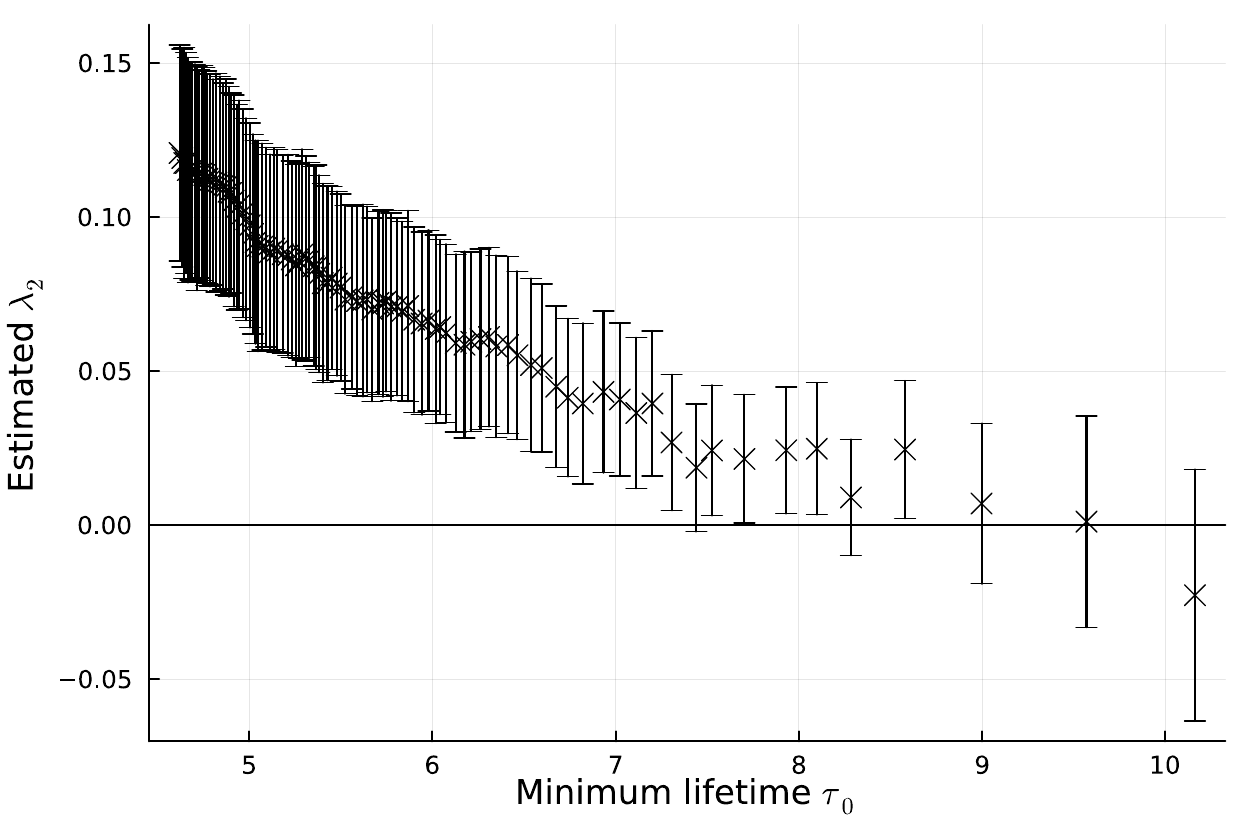}
    \caption[Second Lyapunov exponent as a function of the sampled lifetime]{Estimation of the second Lyapunov exponent $\lambda_2$ for $\sigma_0=0.9$ using only sampled trajectories with a lifetime $\tau > \tau_0$. Each point represents a cut-off at a different percentile (from 1\% to 99\%) of the sampled lifetime distribution, and $\lambda_2$ is estimated in the same way as for Tab. 1 for the selected trajectories. Note that the sample size (originally $N=2000$) strongly decreases for larger $\tau_0$. Error bars indicate one standard deviation of the standard error obtained from bootstrap resampling as in Tab. 1.}
    \label{fig:lambda2-bias}
\end{figure}

\clearpage

\begin{figure}[p]
    \centering
    \includegraphics[width=0.8\textwidth]{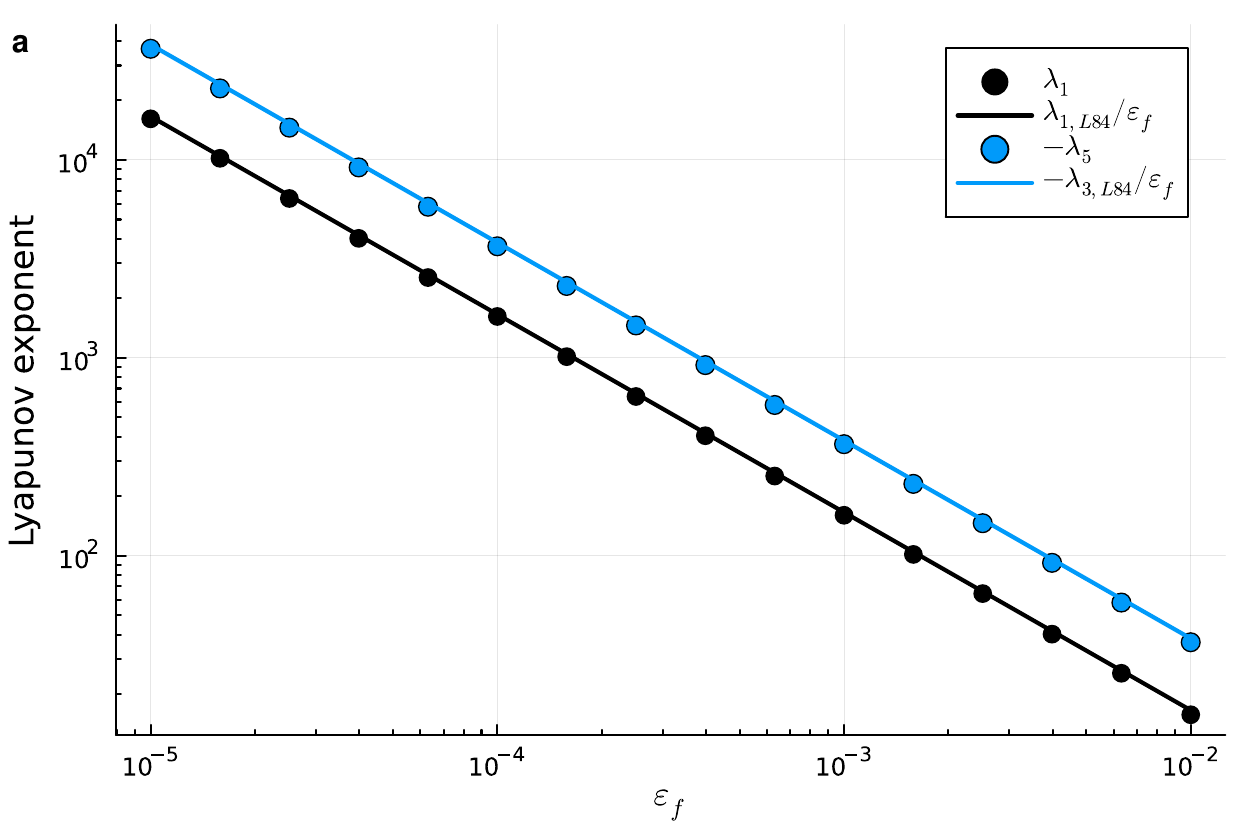}
    \includegraphics[width=0.8\textwidth]{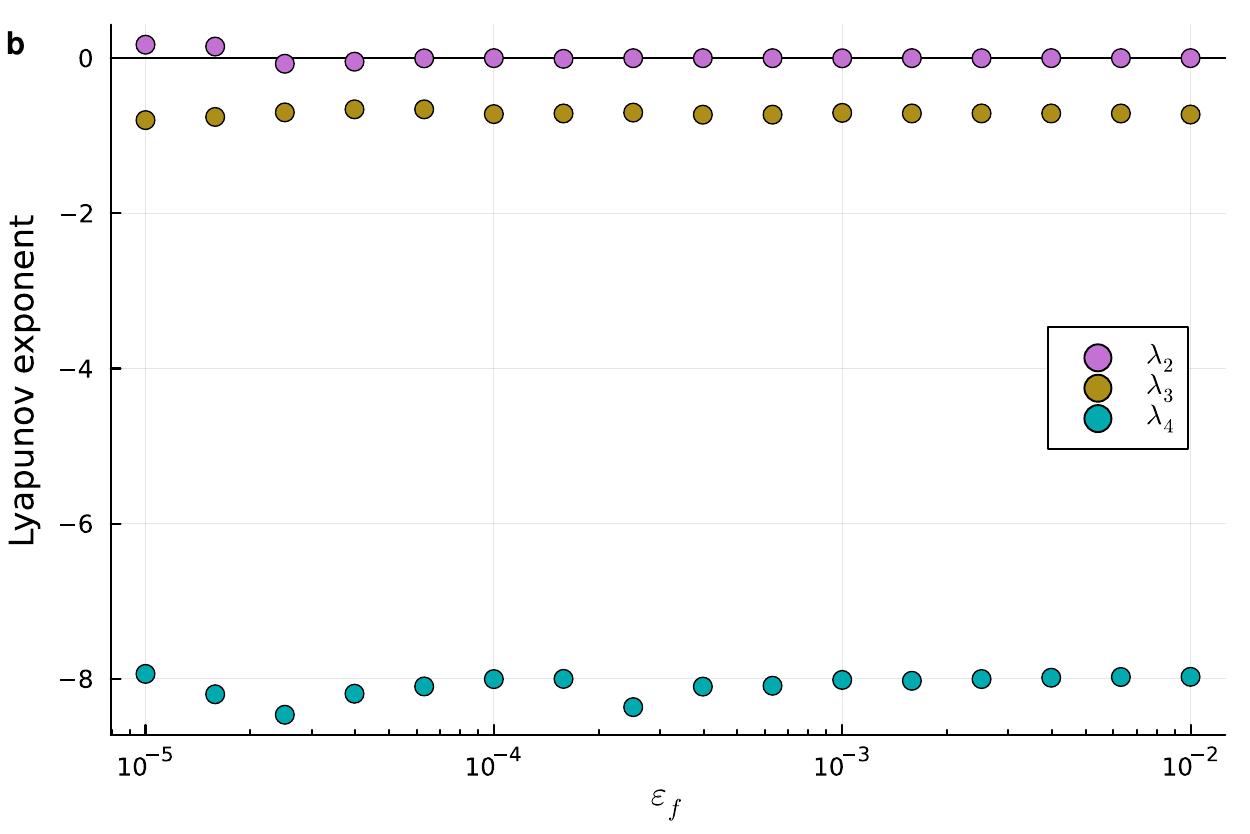}
    \caption[Lyapunov spectrum as a function of $\varepsilon_f$]{Lyapunov spectrum of the coupled L84--Stommel model as a function of the atmosphere--ocean timescale separation $\varepsilon_f$: (a) first and last Lyapunov exponents (dots) and their predicted values from the corresponding LEs of the uncoupled L84 model and $\varepsilon_f$ (lines), (b) second, third and fourth LEs. Note that the $y$-axis is logarithmic in panel a and linear in panel b. As expected, a clear logarithmic dependence on $\varepsilon_f$ can be seen in panel a but not in panel b.}
    \label{sfig:lyapunov-scaling-epsilon}
\end{figure}

\clearpage

\begin{figure}[htbp]
    \centering
    \includegraphics[width=0.8\textwidth]{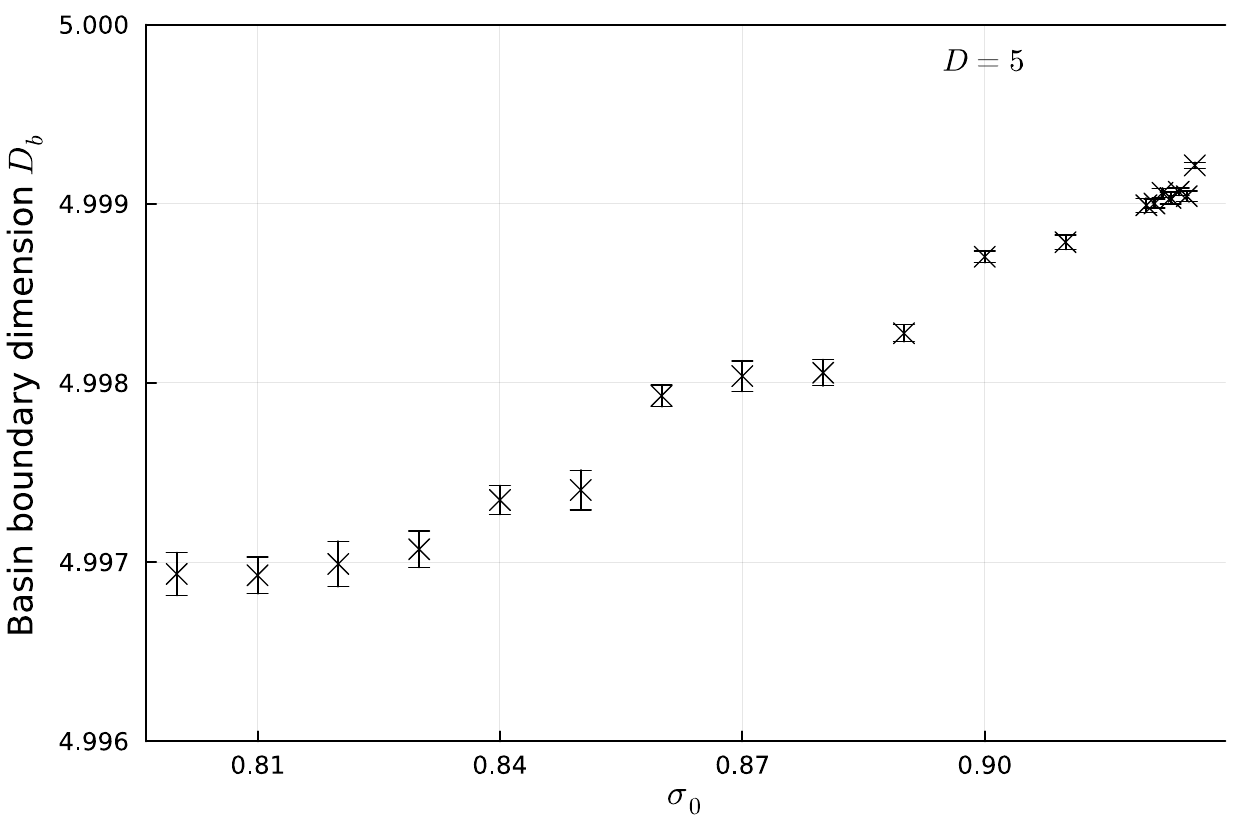}
    \caption[Basin boundary dimension calculated from the saddle lifetime and MLE]{Basin boundary dimension calculated from the saddle lifetime and its MLE following Eq. 6. %\ref{eq:boundary-dim}
    Error bars take into account the uncertainty estimates of the MLEs and the lifetimes. Note that all values are within 1\textperthousand\:of the full phase space dimension $D=5$.}
    \label{sfig:dim-basin-boundary-theo}
\end{figure}

\clearpage

\begin{figure}[htbp]
    \centering
    \includegraphics[width=1\textwidth]{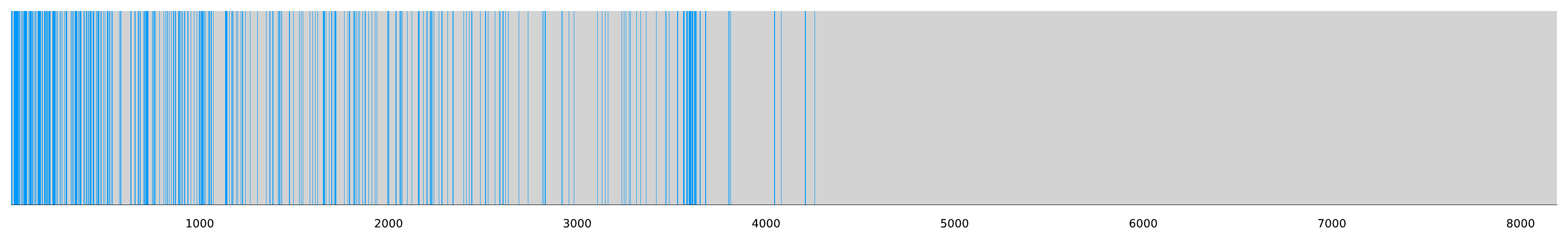}
    \caption[Final state for all $2^{13}$ initial conditions]{Same as Fig. 10a, but for the full ensemble of $2^{13}$ initial conditions. Zoom in to view the individual lines.}
    \label{fig:barcode-zoomout}
\end{figure}

\begin{figure}[htbp]
    \centering
    \includegraphics[width=0.8\textwidth]{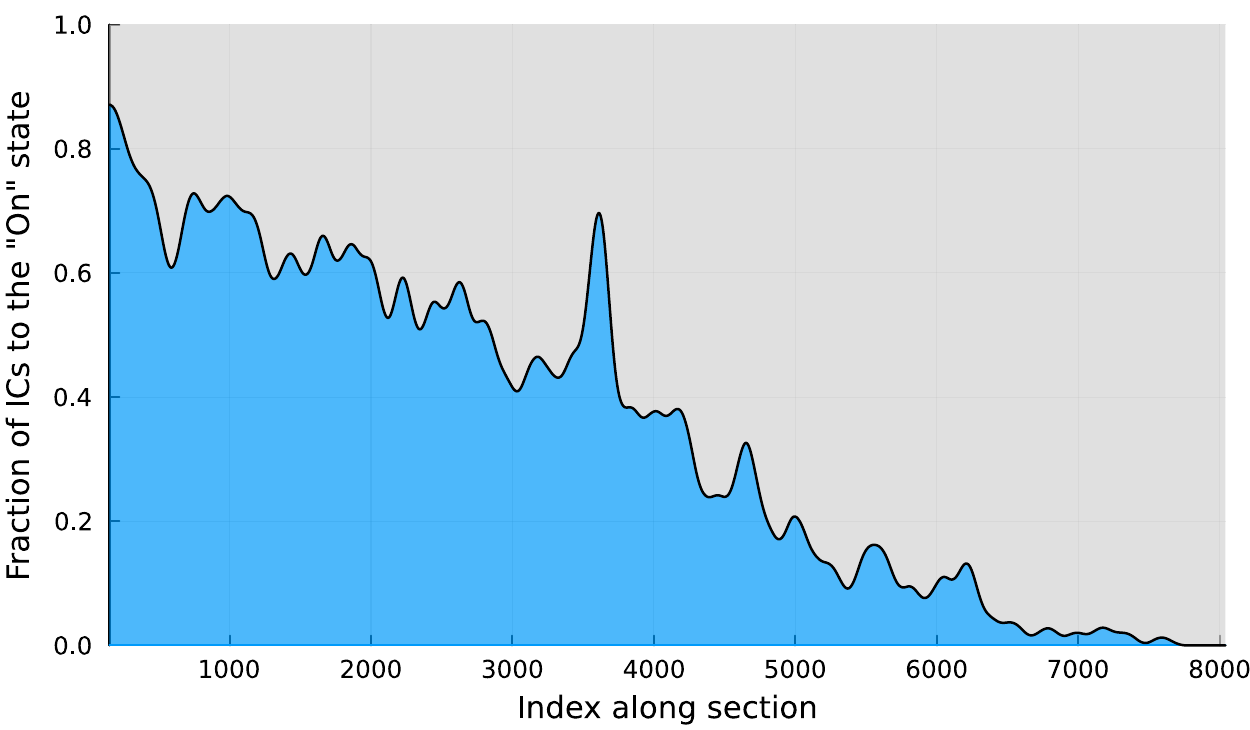}
    \caption[Fraction of initial conditions converging to the two final states]{Fraction of initial conditions coverging to the ``on'' state (black line, blue shading) and to the ``off'' state (grey shading) along the phase space section shown in Fig. \ref{fig:barcode-zoomout}. A Lanczos filter with a bandwidth of 150 neighboring initial conditions is used for smoothing. The fraction of initial conditions converging towards the ``On`` state is clearly, though not monotonously, decreasing as one moves along the phase space section.}
    \label{fig:basin-boundary-fraction}
\end{figure}